\documentclass[aps,pra,reprint,showpacs,showkeys,groupedaddress,nofootinbib,superscriptaddress]{revtex4-1}
\usepackage{hyperref}
\usepackage[T1]{fontenc}
\usepackage[ansinew]{inputenc}
\usepackage{amsmath}
\usepackage{setspace}
\usepackage{psfrag}
\usepackage{mathtools}
\usepackage{amssymb,lineno,amsfonts}
\usepackage{verbatim}
\usepackage{color}
\usepackage{textcomp}
\usepackage[toc,page]{appendix}
\usepackage{graphicx}
\usepackage[caption=false]{subfig}
\usepackage{float}
\usepackage{bm}
\usepackage{dcolumn}
\usepackage{gensymb}
\usepackage{array}
\usepackage{ifpdf}
\usepackage{bbm}
\usepackage{mathrsfs}
\usepackage{upgreek}
\usepackage{epstopdf}
\usepackage{esvect}
\usepackage[usenames,dvipsnames]{xcolor}
\definecolor{med-blue}{RGB}{25,25,112}
\hypersetup{colorlinks, linkcolor={blue},citecolor={blue}, urlcolor={blue}}
\usepackage[english]{babel}
\usepackage[autostyle, english = american]{csquotes}
\usepackage{enumerate}
\usepackage{soul}
\usepackage{amsfonts}
\usepackage{amssymb}
\begin{document}
\title{Multi-component states for trapped spin-1 Bose-Einstein Condensates in the presence of magnetic field}

\author{Projjwal K. Kanjilal}
\email{projjwal.kanjilal@students.iiserpune.ac.in}
 \affiliation{Department of Physics, Indian Institute of Science Education and Research, Dr. Homi Bhabha Road, Pune 411 008, India}
\author{A. Bhattacharyay}
\email{a.bhattacharyay@iiserpune.ac.in}
\affiliation{Department of Physics, Indian Institute of Science Education and Research, Dr. Homi Bhabha Road, Pune 411 008, India}

\date{\today}

\begin{abstract}{ In presence of a magnetic field, multi-component ground states appear in trapped spin-1 Bose-Einstein condensates for both ferromagnetic and anti-ferromagnetic types of spin-spin interaction. We aim to produce an accurate analytical description of the multi-component states which is of fundamental importance. Despite being in the so-called regime of Thomas-Fermi approximation (condensates with large particle number), the scenario of multi-component states is problematic under this approximation due to large variation in densities of the sub-components. We generalize the variational method that we have introduced in the article \cite{kanjilal_variational} to overcome the limitations of T-F approximation. We demonstrate that the variational method is crucial in identifying multi-component ground states. A comparison of the results of the variational method, which is multi-modal by construction, with that of single-mode approximation is also presented in this paper to demonstrate a marked improvement in accuracy over single-mode approximation. We have also looked into the phase transition between the phase-matched and polar state in a trapped condensate using the variational method and have identified substantial change in the phase boundary. The correspondence of the phase diagram of the trapped case with the homogeneous one identifies other limitations of T-F approximation as opposed to the more accurate variational method.}
    



\end{abstract}

\maketitle

\section{Introduction}
The successful experimental realization of Bose-Einstein condensates (BEC) with alkali atoms \cite{1stexps1,1stexps2,1stexps3} inside a magnetic trap spurred renewed interest \cite{earlytheo1,earlytheo2,earlytheo3} in ultracold atomic physics. Soon it attracted a lot of attention from both atomic- and condensed-matter physics communities as it provided an ideal test bed as a quantum simulator \cite{quantumsimu1,quantumsimu2,quantumsimu3,quantumsimu4,quantumsimu5,quantumsimu6} and precision measurements \cite{precisionexp1,precisionexp2,precisionexp3,precisionexp4,precisionexp5,precisionexp6} for its unprecedented experimental control.
\par
Early experiments on BEC \cite{earlyexps1,earlyexps3} were done in magnetic trap which only captures atoms with weak-field-seeking hyperfine states, thus, the magnetic degrees of freedom were frozen in the resulting BEC. Later, with the optical trapping technique this limitation was overcome, and spinor BECs were created with all the hyperfine states of the constituent spin-$f$ atoms ($f$ is an integer) \cite{spinorexp1,spinorexp2,Stamper_Kurn}. The order parameter of such a system has $(2f+1)$ components. Due to the interplay of magnetic field and interatomic interaction, the spinor BEC shows a rich variety of phenomena including spin-textures \cite{spin-textures1,spin_textures_exp}, domain structures \cite{PhysRevLett.81.5718,PhysRevA.85.023601,PhysRevA.84.013619,PhysRevA.96.043603,PhysRevA.60.4857,PhysRevA.78.023632,PhysRevA.80.023602,PhysRevLett.77.3276,2399-6528-2-2-025008,PhysRevA.85.043602,doi:10.1063/1.3243875,doi:10.1142/S0217984917502153,PhysRevA.82.033609,li2018phase,PhysRevA.94.013602,sabbatini2011phase,gautam2014phase,gautam2011phase,PhysRevA.72.023610,Kanjilal_2020,*Kanjilal_corri}, and topological phases \cite{Ueda_2014}. Spinor BEC also attracted a lot of attention due to its complex soliton structures \cite{solitontheo1,solitonexp,solitontheo2,soliton1}, interesting few-body physics in low dimensions \cite{mistakidis_review}.
\par
The role of accurately known density profiles of the multi-component ground state is crucial in dealing with a plethora of interesting phenomena that occur in spinor BEC. As a result, there have been a lot of studies on multi-component ground states \cite{PhysRevLett.81.742,PhysRevA.66.011601,Zhang_2003,PhysRevA.92.023616}. But, most of those analytical studies are based on Thomas-Fermi (T-F) approximation and single-mode approximation (SMA) for the sake of simplicity. However, there exists scope for a wrong interpretation of the ground state structure of multi-component BEC under T-F approximation and similarly wrong estimation of ground state density profiles under SMA which could lead to problems in the presence of closely competing candidate states. 
\par
Both of these problems could be overcome by the method of variational analysis \cite{kanjilal_variational} of ground state profile in a multi-modal manner. One can accurately identify the structure of the tail of density profiles using low-lying oscillator states in a harmonic trap along with the correction to the sub-component density in the central region of the trap. In the present paper, we identify the problems with the T-F and SMA interpretations and compare the results of these methods {\it vis a vis} the more accurate variational method \cite{kanjilal_variational} extended to work for a wide range of magnetic field conditions. 
\par
In this paper, we look at confined spin-1 BEC in the presence of the magnetic field in the absence of any finite temperature \cite{PhysRevA.84.043645,PhysRevA.85.053611,finitetemp} or inter-particle correlation effects \cite{PhysRevA.102.013302}. To get to the ground state one has to solve the Gross-Pitaevskii (GP) equations which govern the dynamics of the three component order parameter (the mean fields $\psi_1$, $\psi_0$ and $\psi_{-1}$). In absence of trapping, solving the GP equations one can get to the phase diagram in $p$, $q$ parameter space \cite{stenger98, KAWAGUCHI2012253,Stamper_Kurn}, where $p$ and $q$ are the linear and quadratic Zeeman terms respectively that capture the contribution coming from the magnetic field.
\par
We consider quasi-one-dimensional $^{87}Rb$ and $^{23}Na$ systems under harmonic trapping. For the $^{87}Rb$, T-F approximation predicts a domain-like structure between the phase-matched (PM) state near the center of the trap followed by the polar state outside. Similarly, for specific choice of Zeeman terms, the T-F approximation predicts the ground state to be a domain structure between the anti-ferromagnetic and ferromagnetic states in $^{23}Na$. Using the generalized multi-modal variational method, we establish that there is no domain structure in the ground states of the above-mentioned cases. The multi-component stationary states at the core of the trap is the ground states in all the cases. 
\par
We also take into account the results of the SMA, which is widely used to capture the physics of spin-oscillation dynamics \cite{KAWAGUCHI2012253,validity_SMA,SMA_physics,SMA1,SMA2}, for a comparison with those of our variational method. We look into an experimentally relevant case where contribution coming from the linear Zeeman term is bypassed by moving to a rotating frame to effectively set $p=0$ \cite{Ganesh1380223}. This is a standard procedure for the application of SMA. However, SMA does not provide a good estimation of the sub-component density profiles for the ground state of the $^{87}Rb$ condensate particularly having large deviations from numerical results across the whole density range. Our multi-modal variational method works quite accurately in comparison to SMA to account for the ground state density profiles. This further emphasizes the merit of the variational method to analytically capture the multi-modal nature of states.
\par
Having shown these results for a quasi one-dimensional trap that standardizes the our variational method to its multi-modal form, in this paper, we embark on identifying the phase boundary between the PM and the polar state for a three dimensional condensate in isotropic harmonic trapping. The phase diagram of between the PM and the polar state is known in the homogeneous (untrapped) case. We show that there arises a significant shift in the phase boundary in the trapped case. However, the phase boundaries for different particle numbers in the trapped case could be collapsed to manifest the correspondence between the trapped and homogeneous systems. The single parameter, dependent on the number of particles in the trapped condensate, that scales the phase boundaries is the value $q_t$ of $q$ at $p=0$ comes out to scale with the particle number as $q_t\sim N^{3/4}$. An equivalent parameter from the mapping of T-F approximated trapped condensate to its homogeneous equivalent results in $q^{T-F}_t\sim N^{2/5}$. This presents a clear distinction of the results of the variational method in comparison to the T-F approximation which could be probed by experiments.
\par
The article is organized in the following way. In section II, we discuss the mean-field theory of the spin-1 trapped BEC in presence of the magnetic field. In section III we focus on the analytical description of the phase-matched state and the anti-ferromagnetic state that become ground states for $^{87}Rb$ and $^{23}Na$, respectively. We start with the T-F results and then we provide a description based on the variational method and compare it with the numerical simulation. We compare the results of SMA with those of the variational method to demonstrate the improvement offered by the latter. Following this, in section IV, we estimate the phase boundary between the PM and the polar state for a similar condensate in isotropic three-dimensional harmonic trap to explore the correspondence to the homogeneous case. This is followed by a discussion on the generality of the treatment presented here and possible directions to explore. 

\section{Mean field theory: GP equation}
The dynamics of the mean fields for a trapped spin-1 BEC in presence of the magnetic field is captured in the Gross-Pitaevskii (GP) equation \cite{PhysRevLett.81.742,doi:10.1143/JPSJ.67.1822,Kanjilal_2020,KAWAGUCHI2012253},
\begin{equation}\label{eq:rev1}
\begin{split}
    i \hbar\dfrac{\partial \psi_m}{\partial t}= \Bigg(-\dfrac{\hbar^2 \nabla^2}{2M} + U(\vec{r})&-pm + qm^2+c_0 n \Bigg)\psi_m\\
    &+c_1 \sum_{m'=-1}^1 \vec{F}.\vec{f}_{mm'} \psi_{m'},
\end{split}
\end{equation} 
 where the first term on the right is the kinetic energy contribution for particles of mass M. The second term is due to the confining potential and the presence of the magnetic field is captured by the linear and quadratic Zeeman terms $p$ and $q$ respectively. We assume a two-body contact interaction that can be decomposed into spin-independent (the term involving $c_0$) and inter-spin interactions (the term with $c_1$). As this is a spin-1 system, the suffix $m$ and $m'$ take the values $1,\ 0$ and, $-1$. 
 \par
 The total density n is defined as,
 \begin{equation}\label{eq:rev2}
     n(r)=\sum^{1}_{m=-1} |\psi_m|^2,
 \end{equation}
 
and the local spin density $\vec{F}$ is
\begin{equation}\label{eq:rev3}
    \vec{F}=\sum^{1}_{i,j=-1} \psi^{*}_i \vec{f}_{ij}\psi_j,
\end{equation}
where the $\vec{f}$ is defined via the spin-1 Pauli matrices \cite{KAWAGUCHI2012253}. The GP equation (Eq.\ref{eq:rev1}) is a set of three coupled non-linear partial differential equations that yields the order parameter $(\psi_1,\psi_0,\psi_{-1})$ as its solution.
\par
The mean-field approximated total energy, \cite{KAWAGUCHI2012253}
\begin{equation}\label{eq:rev4}
    \begin{split}
        E=\int d\vec{r} \sum^1_{m=-1} \psi_m^* &\left( -\dfrac{\hbar^2 \nabla^2}{2M} + U_{trap}(\vec{r}) -pm + qm^2  \right)\psi_m\\
        &\qquad\qquad\qquad+ \dfrac{c_0}{2} n^2+\dfrac{c_1}{2} |\vec{F}|^2,
    \end{split}
\end{equation}
can be compared for different stationary states to get to the ground state of the system.
In this paper, we represent the stationary states as $(\mathbbm{n}_1,\mathbbm{n}_0,\mathbbm{n}_{-1})$, where, $\mathbbm{n}_m$ is the placeholder for the binary notation, 0 or 1. If the sub-components are populated we represent it as 1 and if empty we represent it as 0. In this notation, for example, the ferromagnetic state is represented as $(1,0,0)/(0,0,1)$ where the sub-component corresponding to $m=1$ or $m=-1$ is populated.

To simplify the GP equations further, one can write the mean fields in terms of the density and corresponding phase,
 \begin{equation}\label{eq:rev5}
    \psi_m(\vec{r},t)=\sqrt{n_m(\vec{r})}exp(-\dfrac{i\mu t}{\hbar})exp(-i\theta_m),
 \end{equation}
 where the parameter $\mu$ stands for chemical potential.
 One can get the number and phase dynamics separately \cite{Kanjilal_2020} by using this ansatz in Eq.\ref{eq:rev1},  
 \begin{equation}\label{eq:rev6}
    \dot{n}_0(\vec{r})=-\dfrac{4c_1 n_0 \sqrt{n_1n_{-1}}\sin\theta_r}{\hbar},
 \end{equation}
 \begin{equation}\label{eq:rev7}
    \dot{n}_{\pm1}(\vec{r})=\dfrac{2c_1 n_0 \sqrt{n_1n_{-1}}\sin\theta_r}{\hbar},
 \end{equation}
 \begin{equation}\label{eq:rev8}
    \begin{split}
        \hbar\dot{\theta}_0=\dfrac{1}{\sqrt{n_0(\vec{r})}}&\left(\mathcal{H}-\mu\right)\sqrt{n_0(\vec{r})}\\
        &+c_1\left(n_1+n_{-1}+2 \sqrt{n_{-1}n_1}\cos\theta_r\right),
    \end{split}
 \end{equation}
 \begin{equation}\label{eq:rev9}
    \begin{split}
        \hbar\dot{\theta}_{\pm1}=\dfrac{1}{\sqrt{n_{\pm1}(\vec{r})}}&\left(\mathcal{H}-\mu\right)\sqrt{n_{\pm1}(\vec{r})}\pm c_1\left(n_1-n_{-1}\right)+q\\
        &\quad \mp p+c_1n_0\left(1 +\sqrt{\dfrac{n_{\mp1}(\vec{r})}{n_{\pm1}(\vec{r})}}\cos\theta_r\right),
    \end{split}
 \end{equation}
  where, $\mathcal{H}=-\dfrac{\hbar^2 \nabla^2}{2M} + U(\vec{r})+c_0 n$ and $\theta_r$ is the relative phase which is defined as, $\theta_r=\theta_1+\theta_{-1}-2\theta_0$ \cite{KAWAGUCHI2012253}.
The same ansatz (Eq.\ref{eq:rev5}) makes the energy a function of the sub-component number density and the relative phase,
 \begin{equation}\label{eq:rev10}
    \begin{split}
        E&=\int d\vec{r} e(\vec{r})\\
        &=\int d\vec{r}\Bigg(-\sum^1_{m=-1} \sqrt{n_m(\vec{r})}\dfrac{\hbar^2 \nabla^2}{2M}\sqrt{n_m(\vec{r})}\\ &\qquad+U(\vec{r})n(\vec{r})-p(n_1-n_{-1})+q(n_1+n_{-1})\\
        &\qquad+\dfrac{c_0}{2}n^2(\vec{r})+\dfrac{c_1}{2}\left(n_1-n_{-1}\right)^2\\
        &\qquad+c_1n_0\left[n_1+n_{-1}+2\sqrt{n_1n_{-1}}cos\theta_r\right]\Bigg),
    \end{split} 
 \end{equation}
 where $e(\vec{r})$ is the energy density. Note that, in this article, we are not looking for vortex solutions. Thus, in Eq.\ref{eq:rev6}-\ref{eq:rev10} we have neglected spatial variation of the sub-component phases, assuming that the phases are either constant or varying slowly.
 \par
 For the stationary states, there is no  temporal variation of the  sub-component number densities and the sub-component phases i.e., the left side of the Eq.\ref{eq:rev6}-\ref{eq:rev9} can be equated to zero. From Eq.\ref{eq:rev6}-\ref{eq:rev7}, one can conclude that at least one of the sub-components should be empty to satisfy the equations. Otherwise, the relative phase has to be either 0 or $\pi$ when all the sub-components are populated. Such a stationary state is also known as the phase-matched (PM) state for $\theta_r=0$ and the anti-phase-matched (APM) state for $\theta_r=\pi$.
 The sub-component phase equations (Eq.\ref{eq:rev8}-\ref{eq:rev9}) for a particular stationary state can be solved to get the sub-component number densities and therefore, the total energy. Before going into that, we will rewrite these equations in a non-dimensional form.
\par
We consider the system to be in quasi-one-dimensional harmonic confinement, i.e., the condensate is elongated along the x-axis. This means the trapping frequency along the x-direction is much less than the geometric mean of the trapping frequency along the other two directions i.e., $\omega_x<<\sqrt{\omega_{yz}}$, where $\omega_{yz}=\sqrt{\omega_y\omega_z}$. The number density and the interaction parameters are scaled as \cite{kanjilal_variational},
    \begin{equation}\label{eq:rev11}
  	    c_0=2 \pi l^2_{yz} l_x\lambda_0\hbar\omega_x, \quad c_1=2 \pi l^2_{yz} l_x\lambda_1\hbar\omega_x,
    \end{equation}
    \begin{equation}\label{eq:rev12}
        u_m=2 \pi l^2_{yz} l_x n_m, \quad r=l_x\zeta
    \end{equation}
    where, $l_x^2=\hbar/(m\omega_x)$, $l_{yz}^2=\hbar/(m\omega_{yz})$, and $N$ is the total number of particles in the condensate. As a result, the parameters $\lambda_0$, $\lambda_1$, $\zeta$ and $u_m$ become all dimensionless. 
    \begin{table*}[htp]
	
	\begin{tabular}{|p{1.3cm}|p{4.5cm}|p{7.5cm}|p{3cm}|}
		\hline
		States & Variation of density & Energy density & $\quad$ Restriction\\
		\hline	
		(1,0,0)  $F1$ &$(\lambda_0+\lambda_1)u(\zeta)=\mu'+p'-q'-\zeta^2/2$  &$\frac{\left[ \zeta^2/2-p'+q'\right]\left[\mu'+p'-q'-\zeta^2/2\right]}{(\lambda_0+\lambda_1)}+\frac{\left[\mu'+p'-q'-\zeta^2/2\right]^2}{2(\lambda_0+\lambda_1)}$ &$none$ \\
		(0,1,0)  $P$ &$\lambda_0u(\zeta)=\mu'-\zeta^2/2$   &$\frac{ \zeta^2/2\left[\mu'-\zeta^2/2\right]}{\lambda_0}+\frac{\left[\mu'-\zeta^2/2\right]^2}{2\lambda_0}$ &$none$ \\
		(0,0,1)  $F2$ &$(\lambda_0+\lambda_1)u(\zeta)=\mu'-p'-q'-\zeta^2/2$ &$\frac{ \left[ \zeta^2/2+p'+q'\right]\left[\mu'-p'-q'-\zeta^2/2)\right]}{(\lambda_0+\lambda_1)}+\frac{\left[\mu'-p'-q'-\zeta^2/2\right]^2}{2(\lambda_0+\lambda_1)}$ &$none$\\
		(1,0,1)  $AF$ & $\lambda_0u(\zeta)=\mu'-q'-\zeta^2/2$ and \quad $(u_1-u_{-1})\equiv F_z=\frac{p'}{\lambda_1}$ & $\frac{ \left[ \zeta^2/2+q'\right]\left[\mu'-q'-\zeta^2/2\right]}{\lambda_0}+\frac{\left[\mu'-q'-\zeta^2/2\right]^2}{2\lambda_0}-\frac{p'^2}{2\lambda_1} $ & $none$ \\
		(1,1,1)  $(A)PM$ & $(\lambda_0+\lambda_1)u(\zeta)=k_1-\zeta^2/2$ where, $k_1=\mu'+\frac{(p'^2-q'^2)}{2q'}$ &$\frac{ \zeta^2/2\left[k_1-\zeta^2/2)\right]}{\lambda_0+\lambda_1}+\frac{\lambda_1}{2}\left[\frac{k_1-\zeta^2/2}{\lambda_0+\lambda_1}-\frac{p'^2-q'^2}{2q'\lambda_1}\right]^2+\frac{\lambda_0}{2}\left[\frac{k_1-\zeta^2/2}{\lambda_0+\lambda_1}\right]^2 $ &$PM(|p'|<|q'|)$ $\quad $ $APM \quad (|p'|>|q'|)$ \\
		\hline
	\end{tabular}\\
	\caption{The density and the energy density expressions corresponding to different stationary states at $\lambda_1\neq0$ obtained via T-F approximation are shown here \cite{Kanjilal_2020,*Kanjilal_corri}. All the parameters in this table are in dimensionless form. One can use Eq.\ref{eq:rev11}-\ref{eq:rev12} to convert expressions into dimensional forms.  The energy expressions and the density expressions for PM and APM states are identical. However, PM and APM states are restricted in space where the APM state exists if the absolute value of the linear Zeeman term is higher than that of the quadratic Zeeman term and PM state exists otherwise. }	
    \end{table*}
    \par
	The phase equations can now be written as  (imposing the stationarity condition in Eq.\ref{eq:rev8}-\ref{eq:rev9}),
    \begin{equation}\label{eq:rev13}
	    \begin{split}
	        \bigg\{& -\dfrac{1}{2}\dfrac{d^2}{d\zeta^2}+\dfrac{1}{2}\zeta^2+ \lambda_0 u-\mu'\\
	        &+\lambda_1\left(u_1+u_{-1}+2 \sqrt{u_{-1}u_1}\cos\theta_r\right) \bigg\} \sqrt{u_0}=0,
	    \end{split}
	\end{equation} 
	\begin{equation}\label{eq:rev14}
		\begin{split}
		\bigg\{& -\dfrac{1}{2}\dfrac{d^2}{d\zeta^2}+\dfrac{1}{2}\zeta^2+ \lambda_0 u-\mu' \pm \lambda_1\left(u_1-u_{-1}\right)\\
		&\pm p'+ q'\bigg\}\sqrt{u_{\pm1}}
		+\lambda_1 u_0\left(\sqrt{u_{\pm1}} +\sqrt{u_{\mp1}}\cos\theta_r\right)=0.
		\end{split}
	\end{equation}
    
	where, $\mu'$, $p'$, and $q'$ correspond to the dimensionless forms of chemical potential, the linear and quadratic Zeeman terms respectively. The scaling is done by dividing the parameters with the factor $(\hbar \omega_x)$. The sub-component densities add up to provide the total density of the system, i.e. $u=u_1+u_0+u_{-1}$. 
	\par
	If the kinetic energy contribution is negligible in comparison to interaction terms, then one can use the T-F approximation and solve Eq.\ref{eq:rev13}-\ref{eq:rev14} to get the sub-component number densities hence, the energy density or the total energy for different stationary states. The T-F predicted ground states of our present interest are detailed in Table I.
	
\section{Multi-component stationary states:}
Consider a quasi-one dimensional cigar-shaped harmonic confinement of trapping frequency along the elongated direction $\omega_x=2\pi\times 50 \hspace{0.1cm} Hz$ and the geometric mean of the trapping frequencies along the transverse direction $\omega_{yz}=2\pi\times 1261 \hspace{0.1cm} Hz$.
In the article \cite{kanjilal_variational}, it was shown that for 1-D trapping geometry with the same trapping frequency, the T-F approximation gives reasonably good results in predicting the number density for single component stationary states for $N\geq 500$ in the absence of magnetic field. There is reason to believe that T-F approximation might produce fairly accurate results for multi-component stationary states in presence of a small magnetic field if $N\geq 500$. For the case studies in this paper, the number of condensate particles is fixed at $N=5000$ for which T-F should produce even better results. However, in what follows we will show that even at such a high particle number, T-F results falter and corrections are needed.
\par
Our present focus is on the PM state which is a multi-component state that appears as a ground state for a range of $p$ and $q$ values in condensates with the ferromagnetic spin-spin interaction e.g., $^{87}Rb$. The anti-ferromagnetic state that becomes the ground state for $^{23}Na$ where the spin-spin interaction is of anti-ferromagnetic type \cite{KAWAGUCHI2012253} is also discussed in this paper.
Note that, the quasi-one-dimensional confinement is taken for convenience in numerical simulation. However, the analytic formalism developed and validated is general and can be extended to higher dimensional cases where numerical analysis could be problematic. In the following, we show the results of the T-F study for these two cases, in order to be able to draw a detailed comparison with the beyond T-F results later.

\subsection{PM state: T-F study}
For $^{87}Rb$ condensate with ferromagnetic type interaction, the numerical values of the parameters corresponding to the trap geometry are $l_x=1.53 \hspace{0.1cm}\micro m$, $l_{yz}=0.30 \hspace{0.1cm}\micro m$, 
    $\lambda_0=17.66\times 10^{-3}$ and $\lambda_1=-6.22\times 10^{-4}$.
In this section, to compare the T-F result with the numerical profile as a case study the linear and quadratic Zeeman terms are fixed at $p'=0.01$ and $q'=0.3$. The stationary state that is energetically favorable to be the ground state at these parameter values is the PM state.
Note that, the conclusions, however, are not dependent on a specific choice of parameter values but rather will remain valid for a range of $p'$ and $q'$ values for which the PM state is favorable as the ground state.
\begin{figure*}[htp]
    \centering
	\subfloat[T-F profile of PM state and numerically obtained sub-component number densities\label{subfig-1: sub-component-TF}]{%
		\includegraphics[width=0.40\textwidth]{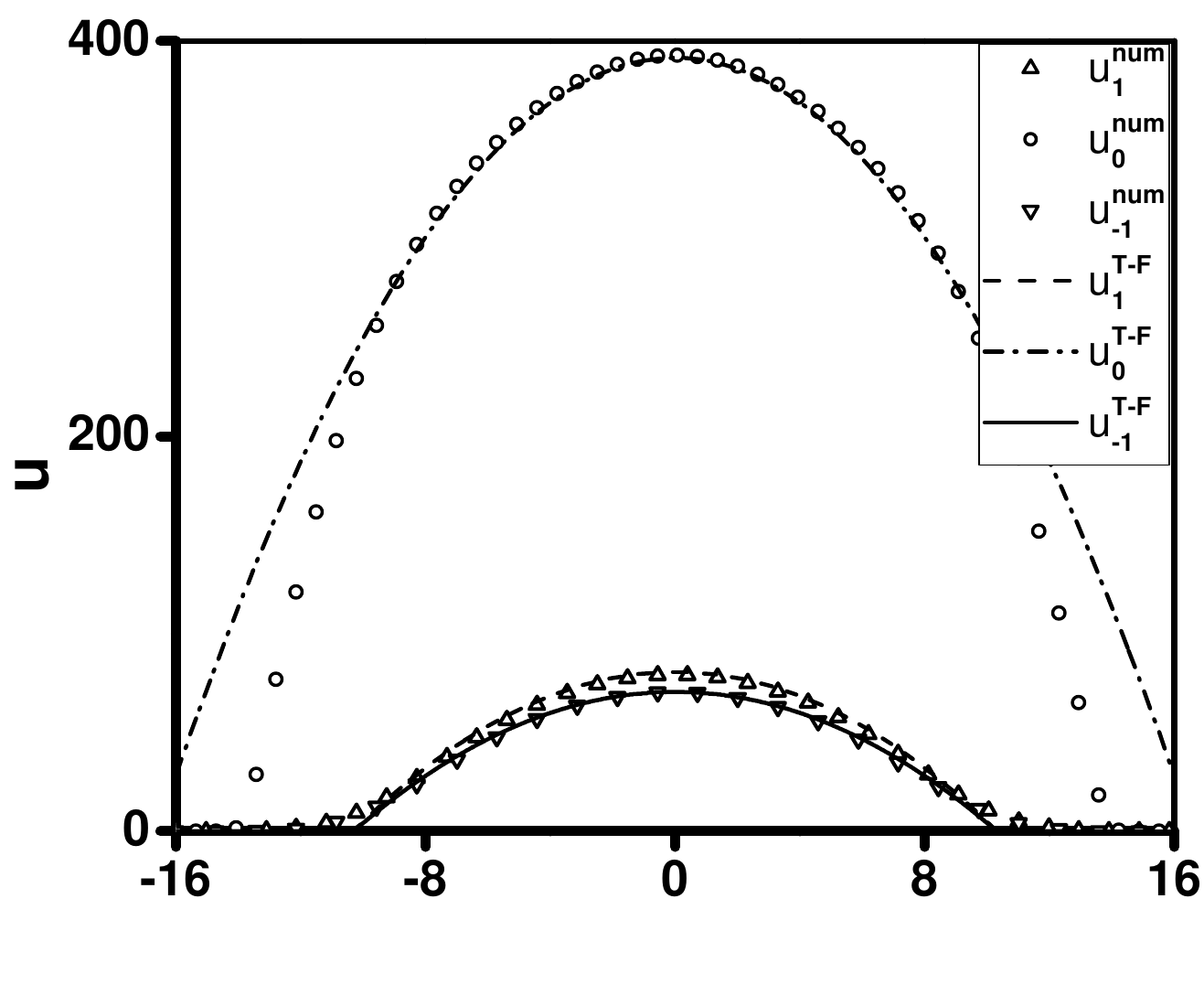}
	}
        \subfloat[T-F predicted domain-like structure and numerically obtained profile of $u_0$ component\label{subfig-2: PM-pol-density}]{%
		\includegraphics[width=0.40\textwidth]{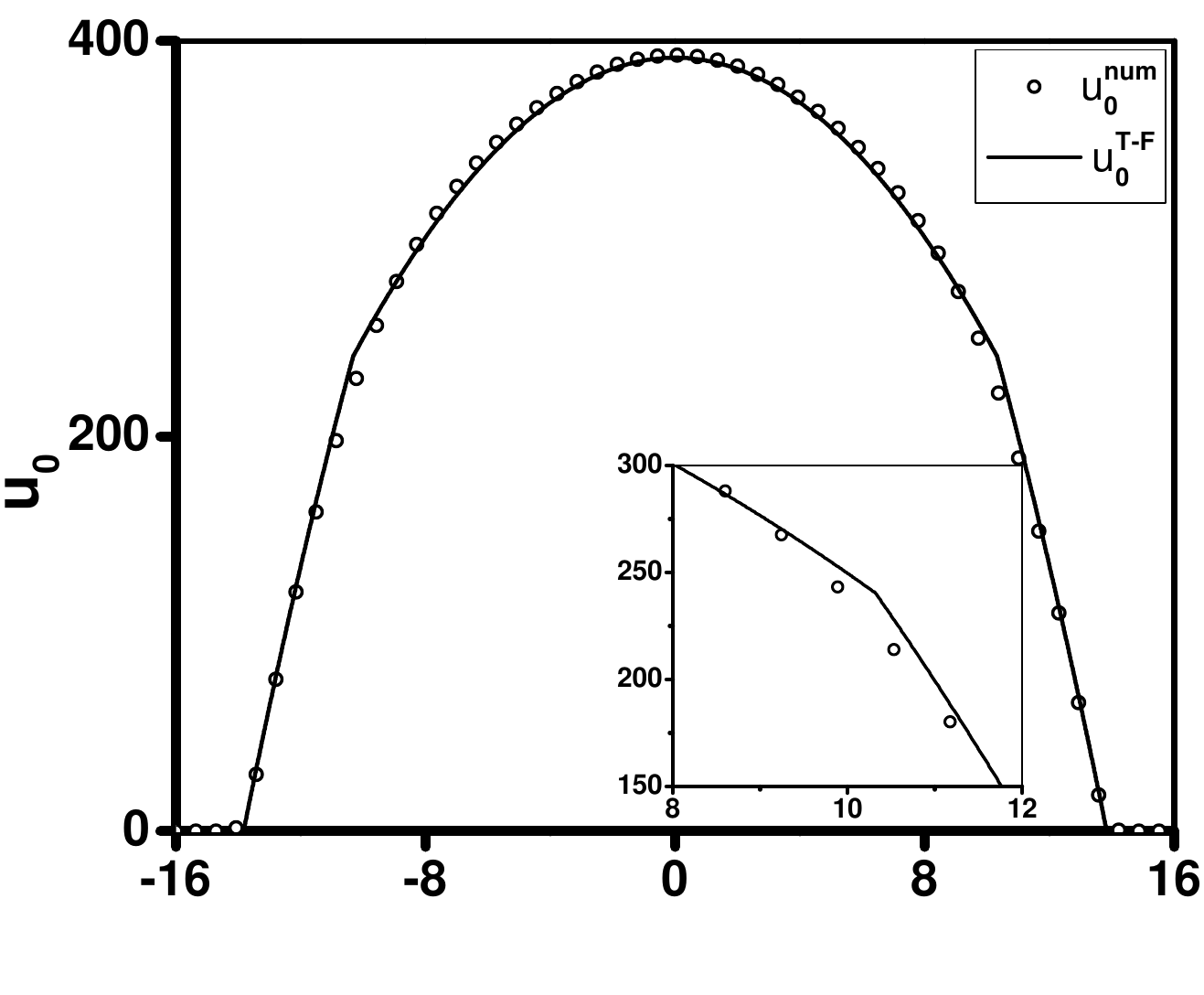}
	}
	\caption{Subfig-(a): The T-F sub-component number densities for PM state is compared with the numerical densities for the parameter value $p'=0.01$ and $q'=0.3$. The T-F approximated $u_0$ expression corresponding to the PM state (shown in the dash-dot line) starts to disagree with that of the numerical simulation beyond $|\zeta|>10.3$, which is the T-F radius of the $u_{\pm1}$ component. Subfig-(b): The T-F prediction of the domain-like situation is plotted where, the $u_0$ component (solid line) follows the T-F expression of the PM state when all the sub-components are populated ($|\zeta|<10.3$ for this case), followed by a polar state-like behavior. Note that discontinuity appears in the $u_0$ (inset) while the numerical result is smooth (bubble markers).}
	\label{fig:number density}
\end{figure*}
\par
The sub-component number densities of the PM state are \cite{Kanjilal_2020,*Kanjilal_corri},
\begin{equation}\label{PMn1}
     u_1^{TF}=\dfrac{(p'+q')^2}{4 q'^2}\left[\dfrac{\mu'+\dfrac{(p'^2-q'^2)}{2 q'}-\dfrac{1}{2} \zeta^2}{\lambda_0+\lambda_1}+\dfrac{q'^2-p'^2}{2 \lambda_1 q'}\right],
 \end{equation}
 \begin{equation}\label{PMn2}
     u_{-1}^{TF}=\dfrac{(p'-q')^2}{4 q'^2}\left[\dfrac{\mu'+\dfrac{(p'^2-q'^2)}{2 q'}-\dfrac{1}{2} \zeta^2}{\lambda_0+\lambda_1}+\dfrac{q'^2-p'^2}{2 \lambda_1 q'}\right],
 \end{equation}
 \begin{equation}\label{PMn0}
     u_{0}^{TF}=\dfrac{(q'^2-p'^2)}{2 q'^2}\left[\dfrac{\mu'+\dfrac{(p'^2-q'^2)}{2 q'}-\dfrac{1}{2} \zeta^2}{\lambda_0+\lambda_1}-\dfrac{q'^2+p'^2}{2 \lambda_1 q'}\right],
 \end{equation}
 which can be compared with the numerically simulated profiles. For the purpose of numerical simulation, imaginary time propagation in the split-step Fourier method \cite{gautam_GPE_solver} is used. The number density profiles as predicted by the T-F are used as initial seeds.     

\par
Comparison of the sub-component densities of T-F profiles with the numerical ones reveals very good agreement for the sub-components $u_1$ and $u_{-1}$. The profile for the $u_0$ component also agrees in the high-density region of the trap Fig.\ref{fig:number density}\subref{subfig-1: sub-component-TF}, but starts to deviate when the other components $u_1$ and $u_{-1}$ vanish which for this parameter values is at around $\zeta^{TF}_{\pm 1}=\pm 10.3$. Note that, $\zeta^{TF}_{\pm 1}$ is the T-F radius of the $u_{\pm 1}$ components. Naturally, one might expect that, beyond the point, $\zeta^{TF}_{\pm 1}$, the summation of the T-F approximated sub-component number densities (i.e., $u_{tot}=u_1+u_0+u_{-1}$) will not agree with the numerical profile as there is a significant mismatch in the $u_0$ component. 
\par
 This is quite natural as the PM state is only valid as long as all the sub-components are populated. However, according to T-F, the $u_{\pm 1}$ components vanish at $\zeta=10.3$, which is the T-F radius of these two sub-components. So, within the T-F picture, it can be said that near the center of the harmonic trap, the PM state is occupied and beyond the T-F radius of $u_{\pm 1}$, the PM state ceases to exist. In this region, only the $u_0$ component is present, which signifies that it is the polar state (see Table I) that occupies the low-density region of the trap.
 \par
 The sub-component number densities for such a construct can be given as,
 \begin{equation}
     u_1^{TF}=\dfrac{(p'+q')^2}{4 q'^2}\left[\dfrac{\mu'+\dfrac{(p'^2-q'^2)}{2 q'}-\dfrac{1}{2} \zeta^2}{\lambda_0+\lambda_1}+\dfrac{q'^2-p'^2}{2 \lambda_1 q'}\right],
 \end{equation}
 \begin{equation}
     u_{-1}^{TF}=\dfrac{(p'-q')^2}{4 q'^2}\left[\dfrac{\mu'+\dfrac{(p'^2-q'^2)}{2 q'}-\dfrac{1}{2} \zeta^2}{\lambda_0+\lambda_1}+\dfrac{q'^2-p'^2}{2 \lambda_1 q'}\right],
 \end{equation}
\begin{equation*}
 u_0^{TF}= 
\begin{dcases}
   \dfrac{(q'^2-p'^2)}{2 q'^2}\Bigg[\dfrac{\mu'+\dfrac{(p'^2-q'^2)}{2 q'}-\dfrac{1}{2} \zeta^2}{\lambda_0+\lambda_1}-\dfrac{q'^2+p'^2}{2 \lambda_1 q'}\Bigg],
   &\\
   \hspace{5cm}\text{if } |\zeta|\leq \zeta_{\pm1}^{TF}\\
   \dfrac{\mu'_{polar}-\dfrac{1}{2}\zeta^2}{\lambda_0},  &\hspace{-2cm}\text{otherwise},
\end{dcases}
\end{equation*}
 where, $\zeta_{\pm1}^{TF}$ is the Thomas-Fermi radius of the $u_{\pm 1}$ component for this 1-D geometry. As can be seen in Fig.\ref{fig:number density}\subref{subfig-2: PM-pol-density} this domain-like explanation works really well as we compare the sub-component density with the numerical $u_0$, but there will be a discontinuity at $|\zeta|=\zeta_{\pm1}^{TF}$. The slope of analytical $u_0$ also changes drastically around this point resulting in a lot of kinetic energy cost.
\par
 If we add the sub-component densities of the PM state given by expressions Eq.\ref{PMn1}-\ref{PMn0} one arrives at the total number density,
\begin{equation}\label{PMntot}
    u_{tot}^{PM}=\dfrac{\mu'+\dfrac{(p^2-q^2)}{2 q}-\dfrac{1}{2} \zeta^2}{\lambda_0+\lambda_1},
\end{equation}
as long as all the sub-components are populated, a necessary condition for the validity of the PM state. As the number density cannot be negative, at $|\zeta|>|\zeta_{\pm 1}^{TF}|$ where the $u_{\pm 1}^{TF}$ goes to zero, the above expression is not valid. Also, the PM state does not exist beyond this point.
\begin{figure}[htp]
	\includegraphics[width=0.40\textwidth]{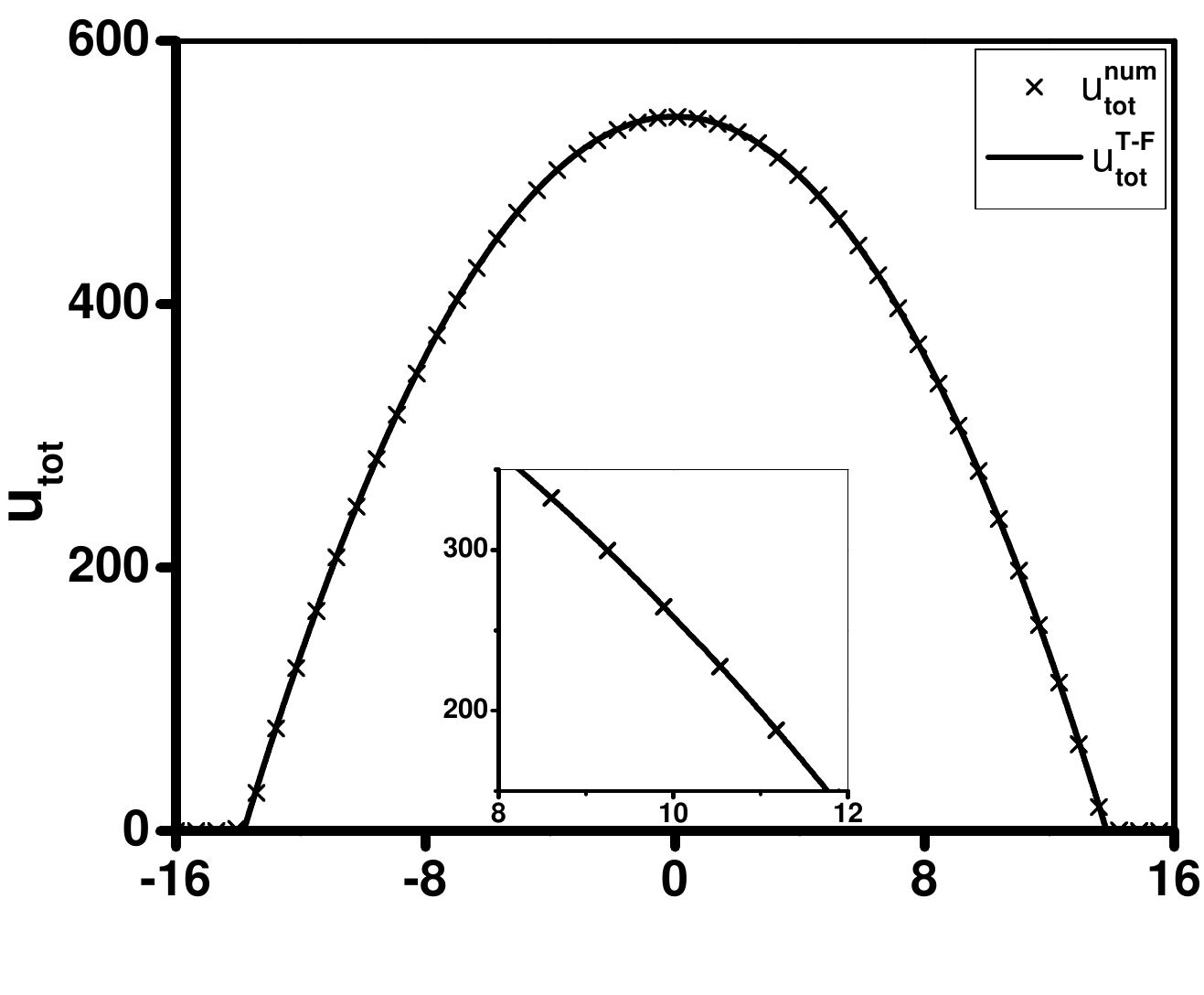}
	
	\caption{The total density obtained from numerical simulation matches quite well with the expression Eq.\ref{PMntot}, which is the T-F expression corresponding to the total density of the PM state. This expression of total number density is only valid as long as all the sub-components are populated. Beyond $|\zeta^{TF}_{\pm1}|$, according to T-F, Eq.\ref{PMntot} describes the total density of the PM state taking negative density contribution from $u_{\pm 1}$ components. (inset) The same expression matches with the numerical total density even beyond the T-F radius $|\zeta^{TF}_{\pm1}|$ which is roughly at $|\zeta|\simeq10.3$.}
	\label{fig:total number density}
\end{figure}
\par
In Fig.\ref{fig:total number density}, we notice that the total number density obtained from the numerical simulation can be aptly described by the analytical expression Eq.\ref{PMntot}.
The element of surprise is that even beyond $|\zeta_{\pm 1}^{TF}|$, where according to T-F the PM state ceases to exist, the numerical profile of total number density agrees with Eq.\ref{PMntot}. Clearly, this cannot be explained via T-F. Beyond $|\zeta_{\pm 1}^{TF}|$, Eq.\ref{PMntot}, according to T-F, would only be valid if negative density contribution from $u_{\pm1}$ components are included, which is unphysical.
\par
The above conclusion about the discontinuity in the $u_0$ component 
 and hence the limitations of T-F approximation can be further strengthened if we compare numerical simulation and T-F prediction for a different spin-spin interaction,
$c_1 \xrightarrow{}5 c_1^{Rb}$ while keeping all other parameter values the same. Experimentally, this is possible via Feshbach resonance \cite{Feshbach}.
\begin{figure}[htp]
	\includegraphics[width=0.40\textwidth]{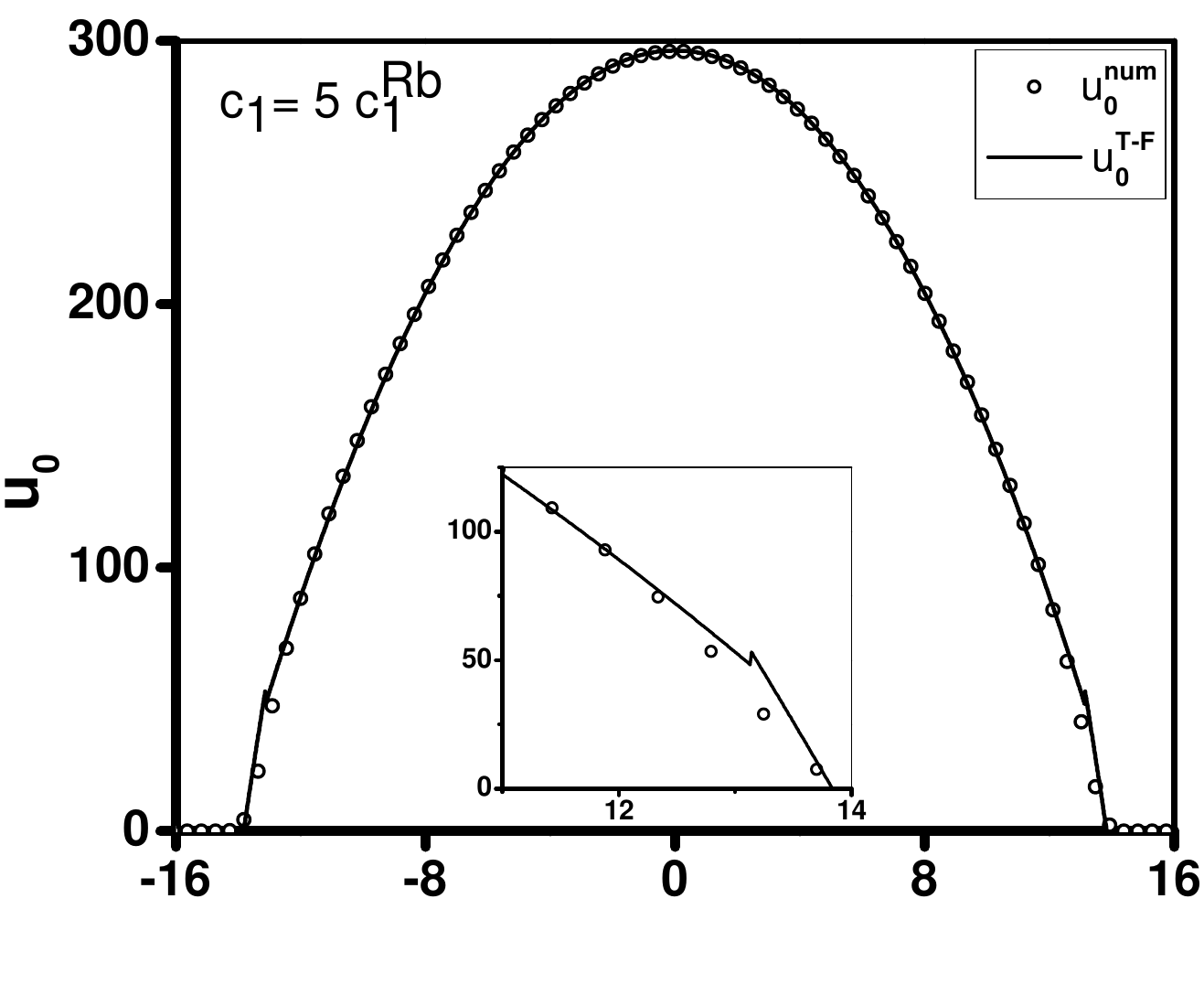}
	\caption{The T-F sub-component number density $u_0$ for PM state for the same the parameter values $p'=0.01$ and $q'=0.3$ and $\lambda_0$ but for a different spin-dependent interaction coefficient $c_1$ (see Eq.\ref{eq:rev4}) which is now 5 times of that of $^{87}Rb$ \cite{KAWAGUCHI2012253}. The PM-like followed by polar-like $u_0$ has a distinct discontinuity at $|\zeta_{\pm 1}^{TF}|=13.13$ (see inset).}
	\label{fig:number density 5c1}
\end{figure}
\par
For this choice of parameter values, the domain-like structure predicted by T-F has a more visible discontinuity (Fig.\ref{fig:number density 5c1}) for the sub-component $u_0$ at
$|\zeta_{\pm 1}^{TF}|=13.13$. The total density for this domain-like structure will also have that same discontinuity. Whereas, the numerical density and the Eq.\ref{PMntot} are in agreement near the T-F radius of $u_{\pm 1}$ and only deviate near the T-F radius of the $u_0$ component itself.
\par
To explain this, one has to remember that in T-F approximation the $u_{\pm 1}$ components sharply go to zero at $|\zeta|=\zeta_{\pm1}^{TF}$ but in reality that cannot be the case. We need to include the kinetic energy term in the analysis. As a result, near this point, the Laplacian terms in Eq.\ref{eq:rev14} cannot be neglected. However, one can ignore the Laplacian term in Eq.\ref{eq:rev13} as the $u_0$ component is still in the high-density region near this point. As the GP equations are coupled, one has to solve the Eq.\ref{eq:rev13}-\ref{eq:rev14} by keeping the Laplacian terms corresponding to $u_{\pm1}$ components. This shows that the T-F approximation is bound to produce an inaccurate description of multi-component stationary states even in the so-called "T-F regime". This warrants the need for an accurate analytical description of the multi-component states appearing for the spin-1 BEC.

\subsection{PM state: The variational method}
In the article \cite{kanjilal_variational}, a multi-modal variational method was introduced which incorporates the kinetic contribution and produces the sub-component number density profiles of a stationary state with great accuracy. The variational method (VM) was developed for a spin-1 BEC in absence of a magnetic field which works really well even for condensates with a small number of particles where the T-F approximation was shown to be no longer valid. The PM state was also analyzed using the multi-modal VM, but for $p=0$ and $q=0$, the sub-components of the PM state follow the same spatial mode, which reduces the complexity significantly.
\begin{figure*}[htp]
	\centering
	\subfloat[The VM approximated sub-component number densities and the numerical profile\label{subfig-1: sub-component-var}]{%
		\includegraphics[width=0.40\textwidth]{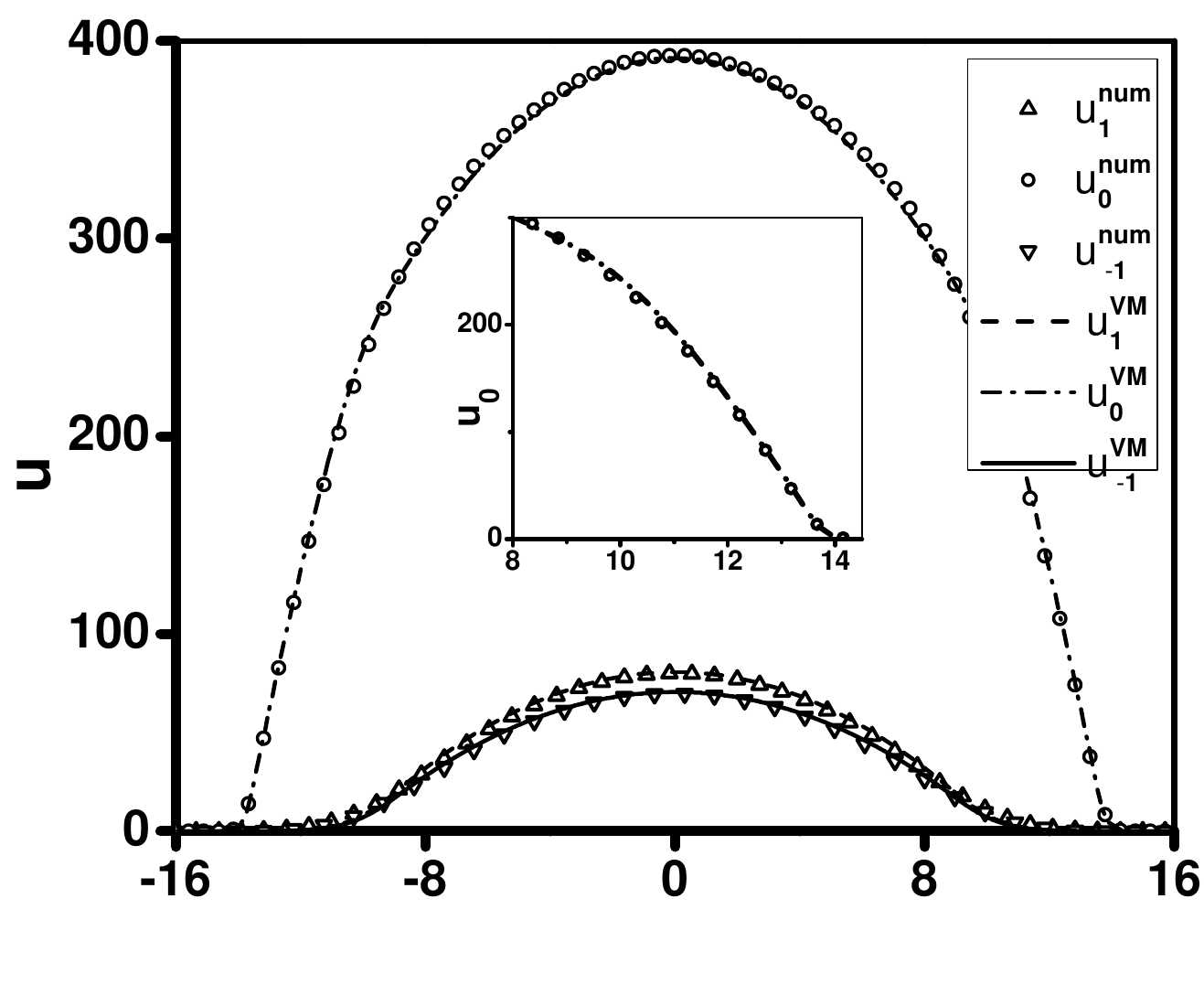}
	}
	\subfloat[The VM approximated and numerically obtained total number density \label{subfig-2: var-vs-num-density}]{%
		\includegraphics[width=0.40\textwidth]{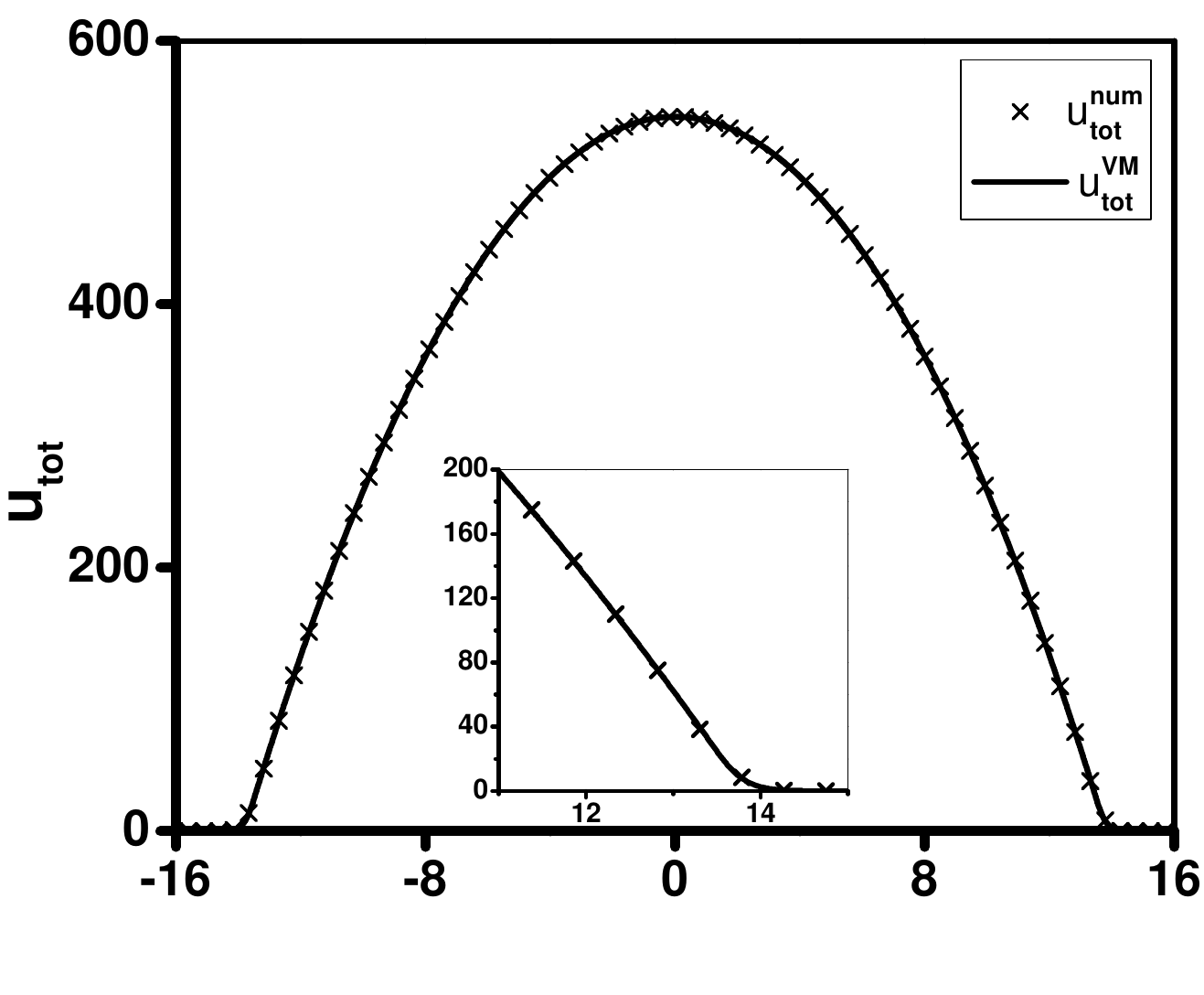}
	}
	\caption{Subfig-(a): The sub-component number density of the PM state obtained via the numerical and variational methods are plotted against the distance from the trap center $\zeta$. The VM produces a very good analytical profile that describes the numerical data quite well even in the tail part of the condensate. The VM rules out any domain-like possibility and analytically estimates the $u_0$ component (shown in the dash-dot line) that is quite accurate in comparison to the numerical (circles) profile (see inset). Subfig-(b): The variationally obtained total number density provides an excellent analytical profile to match the numerical result near the core of the trap as well as in the low-density region, where it gives an analytic estimate of the condensate density which asymptotically goes to zero with the increase of the distance $\zeta$.}
	\label{fig:VM PM}
\end{figure*}
\par
We extend the same procedure here in presence of a magnetic field. If the sub-components do not follow the same spatial variation, the situation is much more complex but the extended VM tackles it with ease. In this section we will briefly discuss the procedure; for a more elaborate description see Appendix A.

\par
For the variational method to work, we need to estimate the sub-component densities in the high-density region from the GP equations (Eq.\ref{eq:rev13}-\ref{eq:rev14}) by neglecting the kinetic energy contribution. This is followed by an assumption of the Gaussian-like tail in the low-density region. So according to this assumption, the sub-component densities can be written as,
\begin{equation}\label{uin}
u^{in}_{\pm 1,0}=g_{\pm 1,0}(\mu',\zeta), \hspace{0.3cm}\text{for}\quad|\zeta|\leq\zeta_{\pm 1,0}^{mat}
\end{equation}
\begin{equation}\label{uout}
\begin{split}
u^{out}_{\pm 1,0}=(a_{\pm 1,0}+c_{\pm 1,0} |\zeta|+d_{\pm 1,0} &\zeta^2)exp\left(-\dfrac{\zeta^2}{b_{\pm 1,0}}\right)\\
&\hspace{0.0cm}\text{for}\quad|\zeta|\geq\zeta_{\pm 1,0}^{mat};
\end{split}
\end{equation}
where $g_{\pm1,0}(\mu',\zeta)$ is the functional form of the sub-component density $u_{\pm 1,0}$ near the center of the trap. In the low-density region, we assume the number density (or the wave function) taking into account the first few lowest harmonic oscillator states.
\par
To determine the unknown coefficients $a_m$, $b_m$, $c_m$, $d_m$ with $m=\pm 1,0$, we impose a smooth matching condition at $|\zeta|=\zeta^{mat}$ where the $\sqrt{u_{\pm1,0}}$ and their first three derivatives become equal for low- and high-density expressions. The third derivative matching also ensures a smooth kinetic energy profile. From these four conditions, one can determine all four coefficients appearing in Eq.\ref{uout}, which now become functions of $\mu'$ and $\zeta^{mat}_{\pm 1,0}$.
If we integrate the sub-component densities and add them, the result would correspond to the total number of condensate particles. From this number conservation equation, the parameter $\mu'$ can be expressed in terms of $\zeta^{mat}_{\pm1,0}$ (see Appendix A for more details). Following this step, the densities and thus, the total energy can be written only as a function of the matching points $\zeta^{mat}_{\pm1,0}$ which are the only variables left. The matching points can be found from the minimization of the total energy. Once the matching points are obtained, we can write the analytical expressions for all the sub-component densities.
\par

Following the numerical evidence that the total number density follows Eq.\ref{PMntot} (see Fig.\ref{fig:total number density}), it is sensible to apply the multi-modal VM for total number density and the $u_{\pm 1}$. The $u_0$ expression can be later found out by subtracting the sum of variational profiles of $u_{\pm 1}$ from the total number density profile, i.e., $u_0^{VM}=u_{tot}^{VM}-u_1^{VM}-u_{-1}^{VM}$. So for simplicity, instead of the $u_0$ component, we will focus on total number density which will also provide a matching point $\zeta^{mat}_{tot}$.
Note that, for the PM state, the spatial mode for the $u_1$ and $u_{-1}$ are equivalent, so the matching point for those two sub-components will be the same i.e., $\zeta^{mat}_1=\zeta^{mat}_{-1}$.
Thus, the minimization of the total energy in the two-dimensional parameter space of $\zeta^{mat}_1$ and $\zeta^{mat}_{tot}$ will determine the energy itself as well as approximate values of these parameters.
\par
For this particular case, we find that $\zeta^{mat}_{\pm1}=8.5$ and $\zeta^{mat}_{tot}=13.43$ minimizes the total energy. These matching points also determine the parameter $\mu'=95.6$ from the number conservation equation.
Thus, number densities can be written in an analytical form as,
\begin{equation}
u_1^{var}= 
\begin{dcases}
1.5173 (53.1007 - 0.5 \zeta^2),\quad \text{if } |\zeta|\leq 8.5\\
\begin{split}
1.5173&(114177.278 - 31226.158 |\zeta|\\
&+ 2184.067 \zeta^2)exp(-0.0824 \zeta^2)
\end{split},&\\
\hspace{4.5cm}\text{otherwise},
\end{dcases}
\end{equation}
\begin{equation}
u_{-1}^{var}= 
\begin{dcases}
1.3278 (53.1007 - 0.5\zeta^2),\quad \text{if } |\zeta|\leq 8.5\\
\begin{split}
1.3278&(114177.278 - 31226.158 |\zeta|\\
&+2184.067 \zeta^2)exp(-0.0824 \zeta^2)
\end{split},&\\
\hspace{4.5cm}\text{otherwise},   
\end{dcases}
\end{equation}
\begin{equation}\label{eq:rev25}
u_{tot}^{var}= 
\begin{dcases}
5.6839 (95.4599 - 0.5 \zeta^2),\quad \text{if } |\zeta|\leq 13.425\\
\begin{split}
5.6839&(4.8883\times10^{25} - 
   7.4283\times10^{24} |\zeta|\\
   &+ 2.8227\times10^{23} \zeta^2)exp(-0.2779 \zeta^2)
\end{split},&\\
\hspace{4.5cm}\text{otherwise},
\end{dcases}
\end{equation}
where the numbers are rounded up to four decimal places. The analytical expressions of the sub-component densities obtained from the VM is in excellent agreement with the numerical profiles (see Fig.\ref{fig:VM PM}\subref{subfig-1: sub-component-var}). Note that, in Eq.\ref{eq:rev25} the coefficients of the total number density might look very large, but for $|\zeta|\geq 13.43$ where the expression is valid, the contribution from the exponential part is so small that the combined contribution asymptotically goes to zero at large distances and matches quite accurately with the numerical profile (see the inset of Fig.\ref{fig:VM PM}\subref{subfig-2: var-vs-num-density})
\par
We have made a case study of the PM state which is a multi-component stationary state that becomes the ground state for a range of linear and quadratic Zeeman strengths. For the purpose of comparison with numerical simulation, we have chosen 1-D harmonic trapping and particular values of $p'$, $q'$, and the number of condensate particles $N$.
\par
Note that the VM is an approximation scheme that works really well in estimating the sub-component number densities (also the mean fields) which produces a very good estimation of the vector order parameter of the spin-1 system. Like other approximate methods, it has some limitations as well. For example, at a large distance from the center of the trap (very large $\zeta$), where the total density $u_{tot}^{VM}$ and $u_{\pm 1}^{VM}$ are very close to zero and can be considered negligible, we find that the total density is slightly lesser than the combined contribution of the $\pm 1$ sub-components hence, making $u^{VM}_0$ slightly negative which is not physical. For this reason, we have taken the contribution up to a large $\zeta$ after which we assume that $u_0$ goes to zero. Thus the kinetic energy contribution is included and considered up to a large distance without discontinuity.

\subsubsection{Comparison with single-mode approximation (SMA)}
\begin{figure*}[htp]
    \centering
	\subfloat[The sub-component number densities obtained from SMA and the numerical profile\label{subfig-1: sub-component-SMA}]{%
		\includegraphics[width=0.40\textwidth]{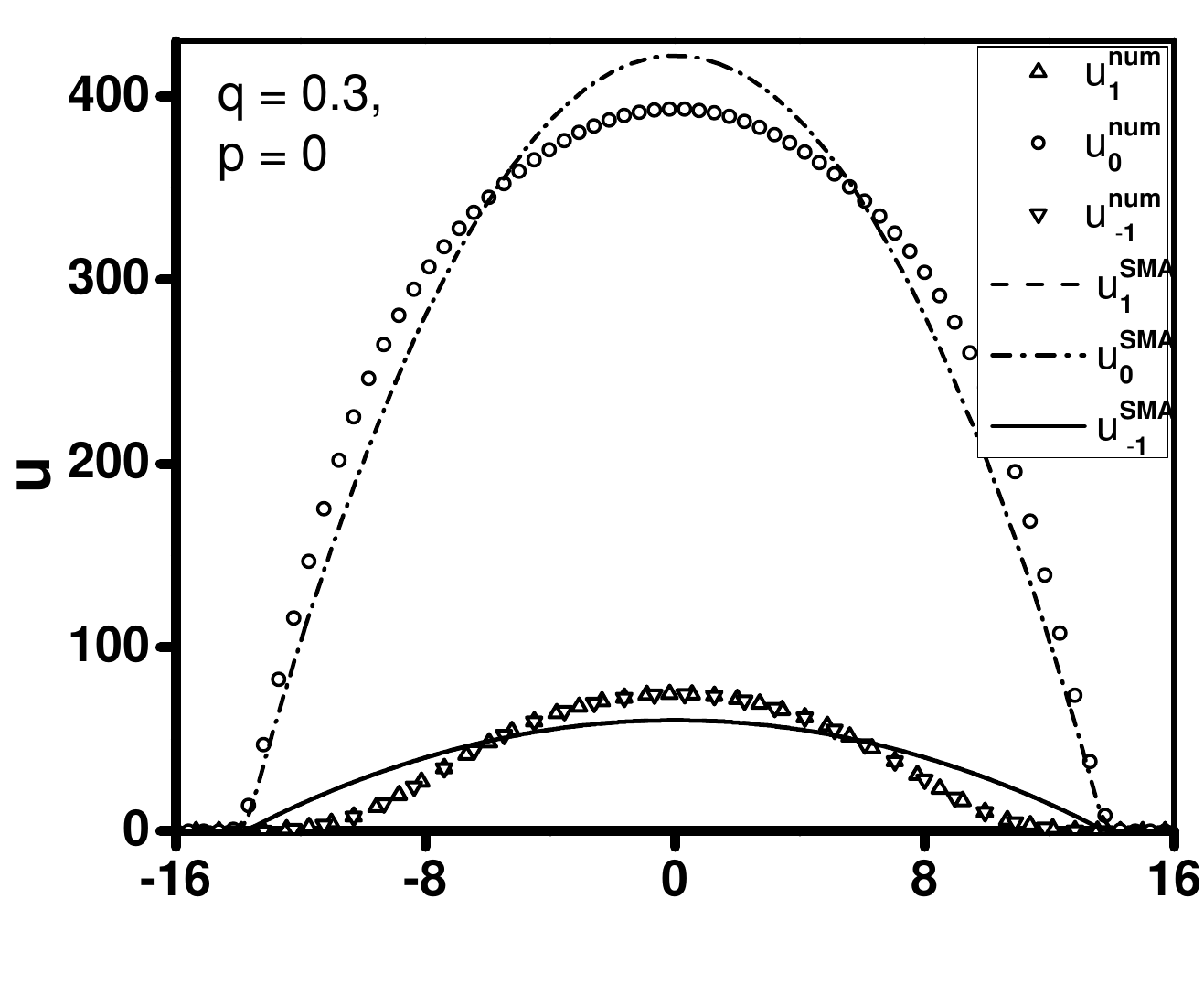}
	}
        \subfloat[The VM approximated and numerically obtained sub-component number densities \label{subfig-2: sub-component varSMA}]{%
		\includegraphics[width=0.40\textwidth]{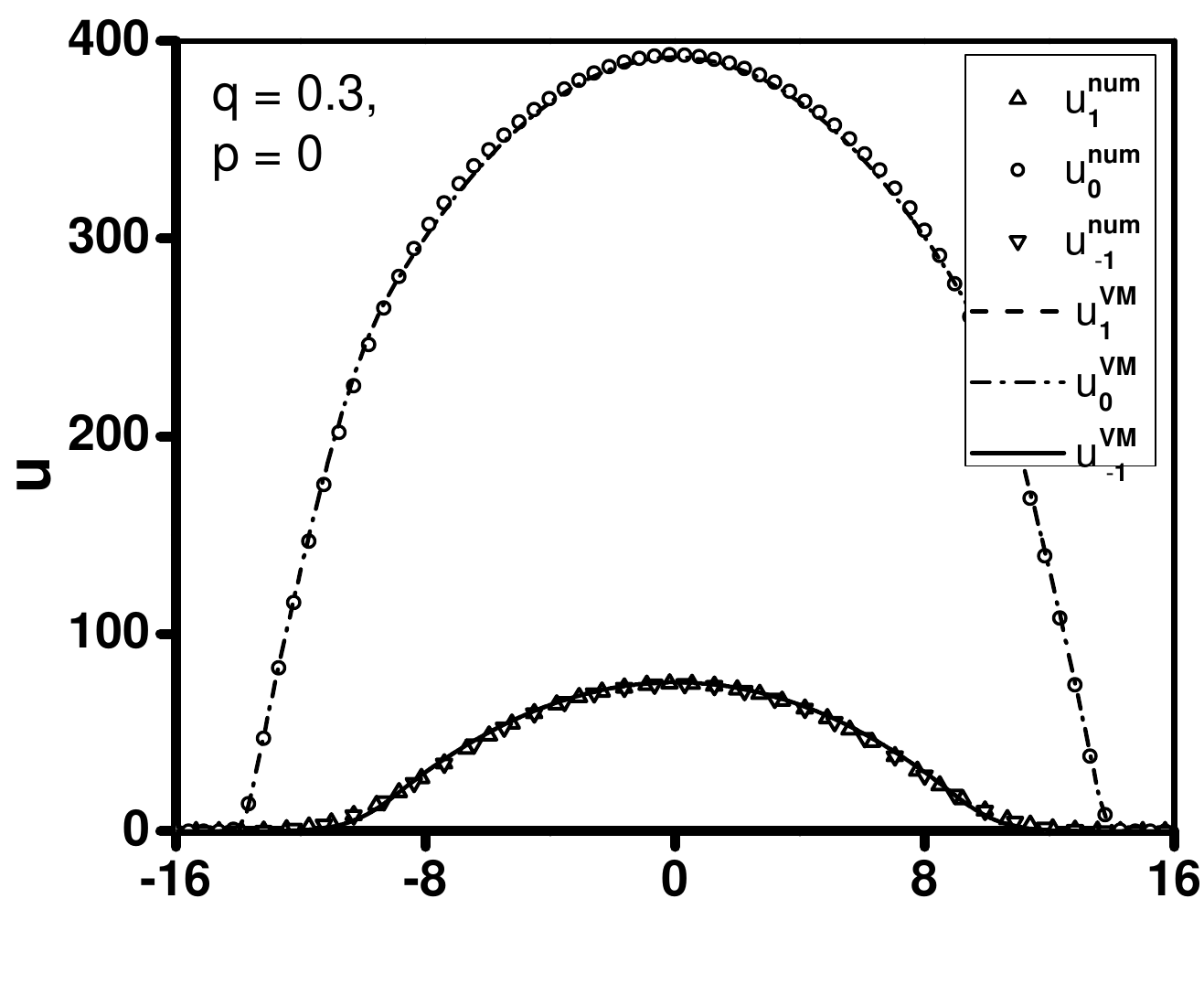}
	}
	\caption{Subfig-(a): The SMA sub-component density profiles are compared with the numerical profile. Though the SMA total number density profile agrees with the numerical total number density distribution (not shown here), the sub-component density profiles are not in agreement. Subfig-(b): The VM estimated sub-component profiles for the case of $p=0$ and $q=0.3$ match quite well with the multi-modal distribution obtained from the numerical estimation.}
	\label{fig:SMA VM}
\end{figure*}
The single-mode approximation is a widely adopted method for the study of spin-oscillation dynamics in spinor condensates. Under SMA, all the sub-components are assumed to follow the same spatial variation \cite{KAWAGUCHI2012253},
\begin{equation}\label{eq:rev26}
 \psi_m(\Vec{r},t)=\sqrt{N}\xi_m(t)\psi_{SMA}(\vec{r})exp\Big(-\dfrac{i\mu t}{\hbar}\Big),
\end{equation}
where, $\psi_{SMA}(\Vec{r})$ is the spatial mode and $\xi_m(t)$ is, in general, a complex quantity that obeys, $\sum^{m=1}_{m=-1}|\xi_m(t)|^2=1$. For 1D harmonic confinement, one can use the same scaling as Eq.\ref{eq:rev11}-\ref{eq:rev12} where,
\begin{equation}
\psi_{SMA}(\zeta)=\sqrt{2\pi l_{yz}^2l_x}\psi_{SMA}(\vec{r}).
\end{equation}
\par
The mode function $\psi_{SMA}(\zeta)$ can be determined by solving,
\begin{equation}\label{eq:rev28}
    \bigg[-\dfrac{1}{2}\dfrac{d^2}{d\zeta^2}+\dfrac{1}{2}\zeta^2+\lambda_0N|\psi_{SMA}(\zeta)|^2\bigg]\psi_{SMA}(\zeta)=\mu' \psi_{SMA}(\zeta),
\end{equation}
subjected to the constraint,
\begin{equation}\label{eq:rev29}
    \int_{0}^{\infty}d\zeta|\psi_{SMA}(\zeta)|^2=1.
\end{equation}
\par
The solution of the equations,
\begin{equation}\label{eq:rev30}
\begin{split}
    i\dfrac{d\xi_{\pm1}}{d\tau}=(\mp &p'+q')\xi_{\pm 1}\\
    &+\Tilde{\lambda}_1\Big[(\rho_{\pm 1}+\rho_0-\rho_{\mp1})\xi_{\pm1}+\xi_0^2\xi^*_{\mp1}\Big],
\end{split}
\end{equation}
\begin{equation}\label{eq:rev31}
    i\dfrac{d\xi_{0}}{d\tau}=\Tilde{\lambda}_1\Big[(\rho_ 1+\rho_1)\xi_0+2\xi_1\xi_{-1}\xi^*_0\Big],
\end{equation}
provides the dynamics of the normalized spinor $\xi_m$, where $\rho_m\equiv |\xi_m(t)|^2$ and $\tau$ is related to time t as, $\tau=\omega_x t$. The effective volume of the system, $V^{eff}\equiv 4\pi l_x l_{yx}^2\Big(\int_{-\infty}^{\infty}d\zeta|\psi_{SMA}(\zeta)|^4\Big)^{-1}$ determines the parameter $\Tilde{\lambda}_1$ \cite{KAWAGUCHI2012253},
\begin{equation}\label{eq:rev32}
\Tilde{\lambda}_1\equiv \dfrac{c_1 N}{\hbar \omega_x V^{eff}}=\dfrac{\lambda_1 N}{2}\int_{-\infty}^{\infty}d\zeta|\psi_{SMA}(\zeta)|^4.
\end{equation}
\par
Rewriting the normalized spinor,
\begin{equation}
\xi_m=\sqrt{\rho_m}\exp\Big(-i\theta_m\Big)\exp\Big(ip'm\tau\Big),
\end{equation}
simplifies Eq.\ref{eq:rev30}-\ref{eq:rev31} as,
\begin{equation}\label{eq:rev34}
\begin{split}
    \dfrac{d\rho_0}{d\tau}=-2\Tilde{\lambda}_1\rho_0\sqrt{(1-\rho_0)^2-f_z^2}\sin\theta_r,
\end{split}
\end{equation}
\begin{equation}\label{eq:rev35}
    \begin{split}
    \dfrac{d\theta_{r}}{d\tau}=-2\Tilde{\lambda}_1\dfrac{(1-2\rho_0)(1-\rho_0)-f_z^2}{\sqrt{(1-\rho_0)^2-f_z^2}}&\cos\theta_r+2q'\\
    &-2\Tilde{\lambda}_1(1-2\rho_0),
\end{split}
\end{equation}
where, $f_z=|\xi_1|^2-|\xi_{-1}|^2$, and $\theta_r$ is the relative phase. From $\rho_0$, one can get to the population fraction in the other two components, i.e., $\rho_{\pm1}=(1-\rho_0\pm f_z)/2$ \cite{KAWAGUCHI2012253}.
\par
We select an experimentally relevant case of $p=0$ and $q=0.3$, which also corresponds to the PM state in the ground state, to compare the SMA with the numerical results. Numerical solution of Eq.\ref{eq:rev28} estimates the mode function which is the same for all the spin components under SMA. Following that, the stationarity condition can be employed in Eq.\ref{eq:rev34}-\ref{eq:rev35} to find the population fraction for different sub-components. When $f_z=0$, for PM state ($\theta_r=0$), we find $\Tilde{\lambda}_1\simeq-0.2703$, $\rho_0\simeq0.777$ and $\rho_{\pm 1}\simeq0.111$. This population fractions determine the sub-component densities as, $u_m^{SMA}=N\rho_m |\psi_{SMA}(\zeta)|^2$.
\par
The total density profile obtained from SMA ($N|\psi_{SMA}(\zeta)|^2$) is in good agreement with the numerically obtained total number density. This is important for an accurate determination of $\Tilde{\lambda}_1$ (Eq.\ref{eq:rev32}). Still, the sub-component density profiles as obtained from SMA do not agree with the numerically obtained profiles (see Fig.\ref{fig:SMA VM}\subref{subfig-1: sub-component-SMA}). Note that, it is well-known that SMA is not exact even in the ground state for the PM state, which is also known as the broken-axisymmetry phase \cite{KAWAGUCHI2012253}. The inaccuracy of SMA further emphasizes the fact that the sub-components do not follow a single spatial mode for the PM state. Thus, a multi-modal analysis is required. In Fig.\ref{fig:SMA VM}\subref{subfig-2: sub-component varSMA} we demonstrate that the sub-component density distribution obtained from the multi-modal VM is in excellent agreement with the numerical simulation for this experimentally relevant case.
\par
In the next section, we will do a brief case study on the anti-ferromagnetic state which is the other possible multi-component stationary state that becomes the ground state for $^{23}Na$.

\subsection{Anti-ferromagnetic state}
For the $^{23}Na$-condensate, we set the same trapping frequencies corresponding to 1-D confinement as mentioned earlier for the ferromagnetic type condensate. The oscillator length in elongated direction $l_x=2.97 \hspace{0.1cm}\micro m$ and in the transverse direction $l_{yz}=0.59\hspace{0.1cm} \micro m$.
Note that, although we consider the same trapping geometry, the oscillator length scale for $^{23}Na$ and $^{87}Rb$ condensates are different due to the different masses of the species. The spin-independent and spin-spin interaction parameters are $\lambda_0=46.16\times 10^{-3}$ and $\lambda_1=7.43\times 10^{-4}$ corresponding to the values given in \cite{KAWAGUCHI2012253}. The positive spin-spin interaction coefficient signifies the anti-ferromagnetic type of spin interaction for $^{23}Na$ condensate. For a range of linear and quadratic Zeeman terms, the anti-ferromagnetic (AF) state is found to be favorable as the ground state but for the purpose of numerical study, we will focus on the case where $p'=0.2$ and $q'=-0.5$.
\par
As long as the $u_1$ and $u_{-1}$ sub-components are non-zero, the T-F approximation gives an estimation of the total as well as sub-component number densities (see Table 1),
\begin{equation}
    u^{TF}_1=\dfrac{\mu'-q'-\zeta^2/2}{2\lambda_0}+\dfrac{p'}{2\lambda_1},
\end{equation}
\begin{equation}
    u^{TF}_{-1}=\dfrac{\mu'-q'-\zeta^2/2}{2\lambda_0}-\dfrac{p'}{2\lambda_1}.
\end{equation}
\par
As the chosen value of $p'$ is positive (also, $\lambda_1>0$ ), the $u_{-1}$ component goes to zero much faster than the other one. So beyond the T-F radius of the $u_{-1}$ component, the AF state ceases to exist but the sole presence of the  $u_1$ component signifies the ferromagnetic state. So according to T-F, the situation is domain-like with the AF state at the center of the trap followed by the ferromagnetic state outside ( for $|\zeta|>\zeta^{TF}_{-1}$).
\begin{figure}[htp]
    \centering
	\subfloat{%
		\includegraphics[width=0.40\textwidth]{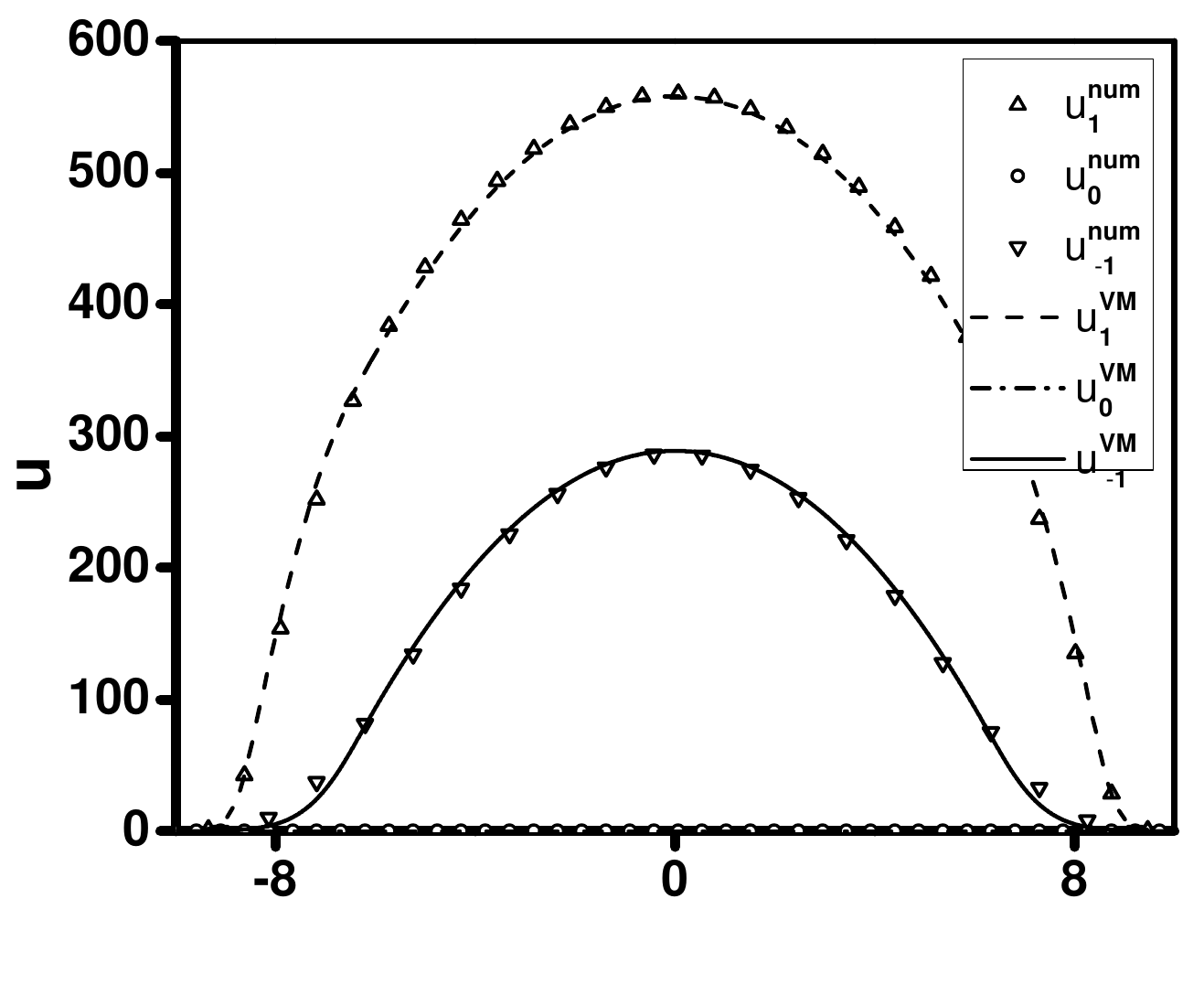}
	}
	\caption{Sub-component density expressions obtained via VM are compared with the numerical profile for the anti-ferromagnetic state with the Zeeman terms fixed at $p'=0.2$ and $q'=-0.5$.}
	\label{fig:VM_anti-ferro}
\end{figure} 
\par
Just like the PM state that we have seen earlier, the numerical simulation does not vindicate the domain-like prediction. Rather the AF state is found to be present for all values of $\zeta$. Following the same procedure as described for the PM state one can apply the VM for the total density and the sub-component density $u_{-1}$. The $u_0$ component is empty and there will be two matching points $\zeta_{tot}^{mat}$ and $\zeta_{-1}^{mat}$ which are to be found out from the minimization of the total energy. For the previously mentioned $p'$ and $q'$ values, we get these two matching points to be, $\zeta_{tot}^{mat}=8.36$ and $\zeta_{-1}^{mat}=6.08$. These also produce the analytical formulae of the total density and the $u_{-1}$ components. The density expression for $u_{1}$ component can be obtained by subtracting the other sub-component density from the total density. 
\par
The VM shows that the domain-like situation that the T-F predicts is incorrect and justifies the fact that the kinetic energy terms cannot be neglected near the T-F radius for the sub-component which is of smaller density, in this case, the $u_{-1}$ component. The VM also produces a low-density expression of the $u_{-1}$ component which has a small but non-zero presence beyond $u_{-1}^{TF}$. Thus, it is only the AF state that is present for all regions of space. Moreover, the analytic 
 number density expressions obtained from VM corresponding to each sub-component are in fair agreement with the numerically obtained profiles (see Fig.\ref{fig:VM_anti-ferro}). 

\section{Phase transition between PM and polar states under confinement}
In the previous section, we have shown that the results of T-F approximation and the SMA are inaccurate for the multi-component ground states under 1-D harmonic confinement. The variational method, on the other hand, analytically obtains the correct profile of the condensate ground states (PM state) as opposed to the T-F approximated domains of PM and polar states. 
\par
Using this variational method one can estimate the phase transitions between different ground states of a trapped spin-1 BEC, especially when the multi-component states are involved. We will choose the case of spin-1 BEC with the ferromagnetic type of spin-spin interaction inside a 3-dimensional (3-D) isotropic harmonic confinement. While the numerical simulation is costlier in higher dimensions, the VM can be implemented to analytically obtain the phase boundaries involving multi-component states.
\par
\begin{figure}[htp]
    \centering
	\subfloat{%
		\includegraphics[width=0.40\textwidth]{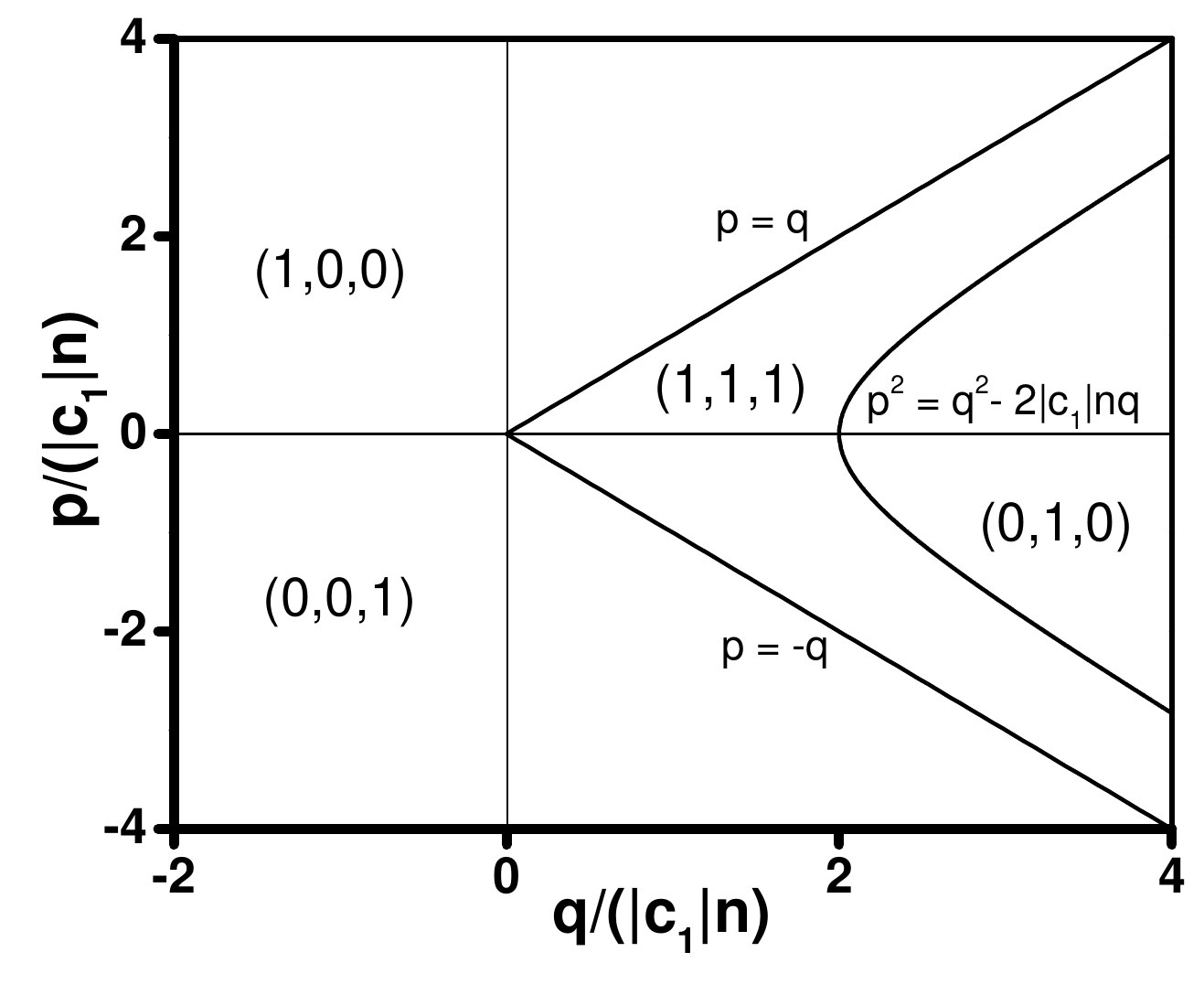}
	}
	\caption{The phase diagram of the spin-1 condensate with a ferromagnetic type of spin-spin interaction ($c_1<0$) in the absence of any confinement. The ferromagnetic, PM, and polar states are favorable to become the ground states depending on the linear and quadratic Zeeman terms, $p$ and $q$. The number density $n$, being a constant over space in the absence of any confinement, can be used to scale the $p$ and $q$ axes. In this scaling, the phase diagram becomes universal in the sense that this diagram does not vary if the number density changes.}
	\label{fig:homogeneous}
\end{figure}
\begin{figure*}[htp]
    \centering
    \subfloat[The energy difference of PM and polar state in absence of trapping.\label{subfig-1: hom_energy}]{%
		\includegraphics[width=0.40\textwidth]{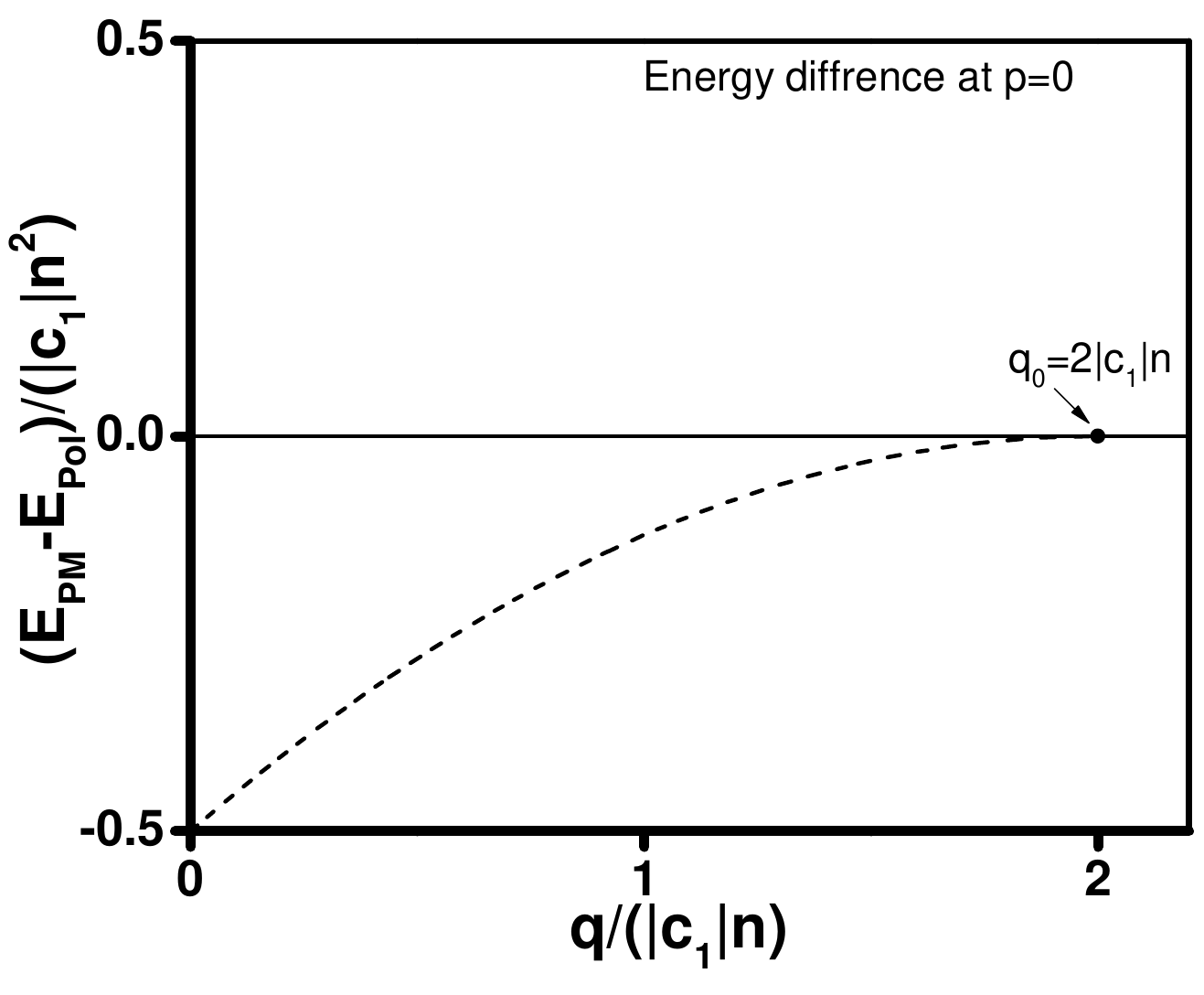}
	}	
    \subfloat[The VM approximated energy difference of the PM and polar state under 3-D isotropic harmonic confinement.\label{subfig-2: VM_energy}]{%
		\includegraphics[width=0.40\textwidth]{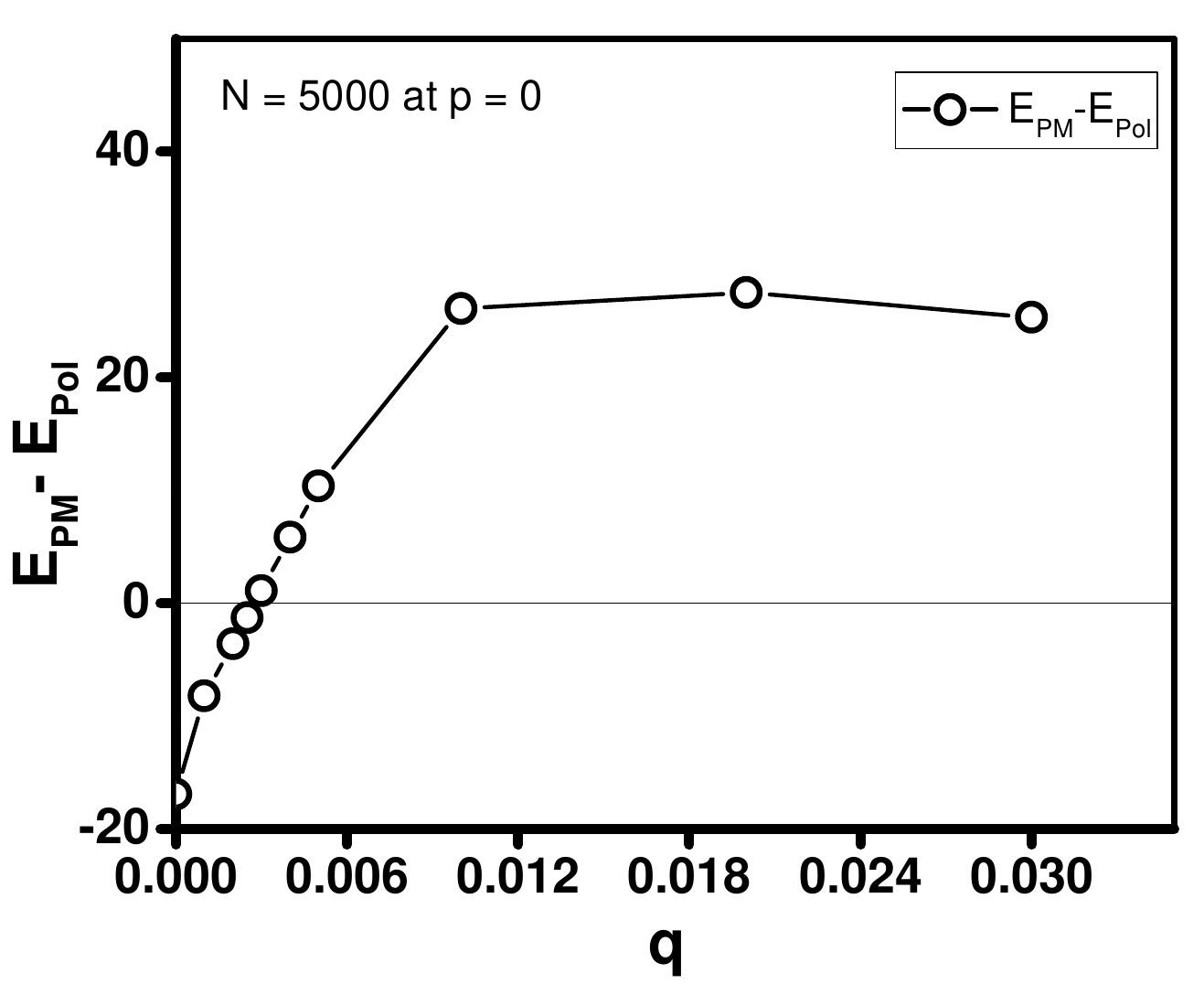}
	}
        
	\caption{Subfig-(a): The energy difference of the PM and the polar state scaled with the constant number density is plotted against the variation of $q$ at $p=0$ for condensates in the absence of any trapping. At $q\approx 0$, the energy corresponding to the PM state is lower than that of the polar state, making the PM state favorable to become the ground state. As $q$ increases, the energy difference reduces, and at the transition point ($q=2|c_1|n$ at $p=0$), the PM state energy becomes equal to that of the polar state. At this point, the sub-component density $n_{\pm 1}$ vanishes. Hence beyond this point, the PM state does not exist. Subfig-(b): The VM estimated energy difference between the PM and polar state under the 3-D harmonic trapping for 5000 condensate particles and at $p=0$. The total energy of the PM state is lower than that of the polar state for small values of the quadratic Zeeman term. The energy difference in the trapped situation indicates that the phase transition happens at $q\simeq 0.0027$, which is much lower than $q\simeq 0.0377$, beyond which the PM state ceases to exist.}
	\label{fig:energy_PM_pol}
 \end{figure*}

In the absence of confinement (hence, a constant number density), the phase diagram for the homogeneous spin-1 condensate is well known. When the spin-spin interaction is of ferromagnetic type ($c_1<0$), the phase diagram of the homogeneous condensate, Fig.\ref{fig:homogeneous} \cite{KAWAGUCHI2012253}, shows different stationary states favorable as the ground state in certain regions of the (q,p) parameter space. For the negative quadratic Zeeman term, the ferromagnetic states are the ground states, where one of them is favorable depending on the sign of the linear Zeeman term. In the positive half, if the quadratic Zeeman term is greater than the absolute value of the linear term (i.e., $q\geq|p|$) then the PM state becomes the ground state as long as $p^2\geq q^2-2 |c_1|nq$ is satisfied, followed by the polar state in the remaining part of the (q,p) parameter space. For the homogeneous condensate, the PM-polar phase transition occurs at $p^2=q^2-2|c_1|nq$. At this point, the energies of the two states become equal and beyond this point the PM state is non-existent. 
  
\par
For an isotropic 3-D confinement of trapping frequency $\omega$, we scale the number density and the interaction parameters as,
\begin{equation}\label{eq:rev38}
    c_0=\dfrac{4 \pi}{3} l_{osc}^3\lambda_0\hbar\omega_x, \quad c_1=\dfrac{4 \pi}{3} l_{osc}^3\lambda_1\hbar\omega_x,
\end{equation}
\begin{equation}\label{eq:rev39}
    u_m=\dfrac{4 \pi}{3} l_{osc}^3 n_m, \quad r=l_{osc}\zeta
\end{equation}
where, $l_{osc}^2=\hbar/(m\omega)$ is the oscillator length scale for this geometry. For this choice of scaling, the phase equations would assume a similar structure as Eq.\ref{eq:rev13}-\ref{eq:rev14} with the Laplacian term to be replaced with $-\dfrac{1}{2}\dfrac{1}{\zeta^2}\dfrac{d}{d\zeta}(\zeta^2\dfrac{d}{d\zeta})$, the radial part (of the Laplacian) in the spherical polar coordinate (Appendix B).
\par
\begin{figure*}[htp]
    \centering
	\subfloat[Phase boundaries of the PM-polar phase transition for different number of particles N.\label{subfig-1: trapped_phase_daigram}]{%
		\includegraphics[width=0.40\textwidth]{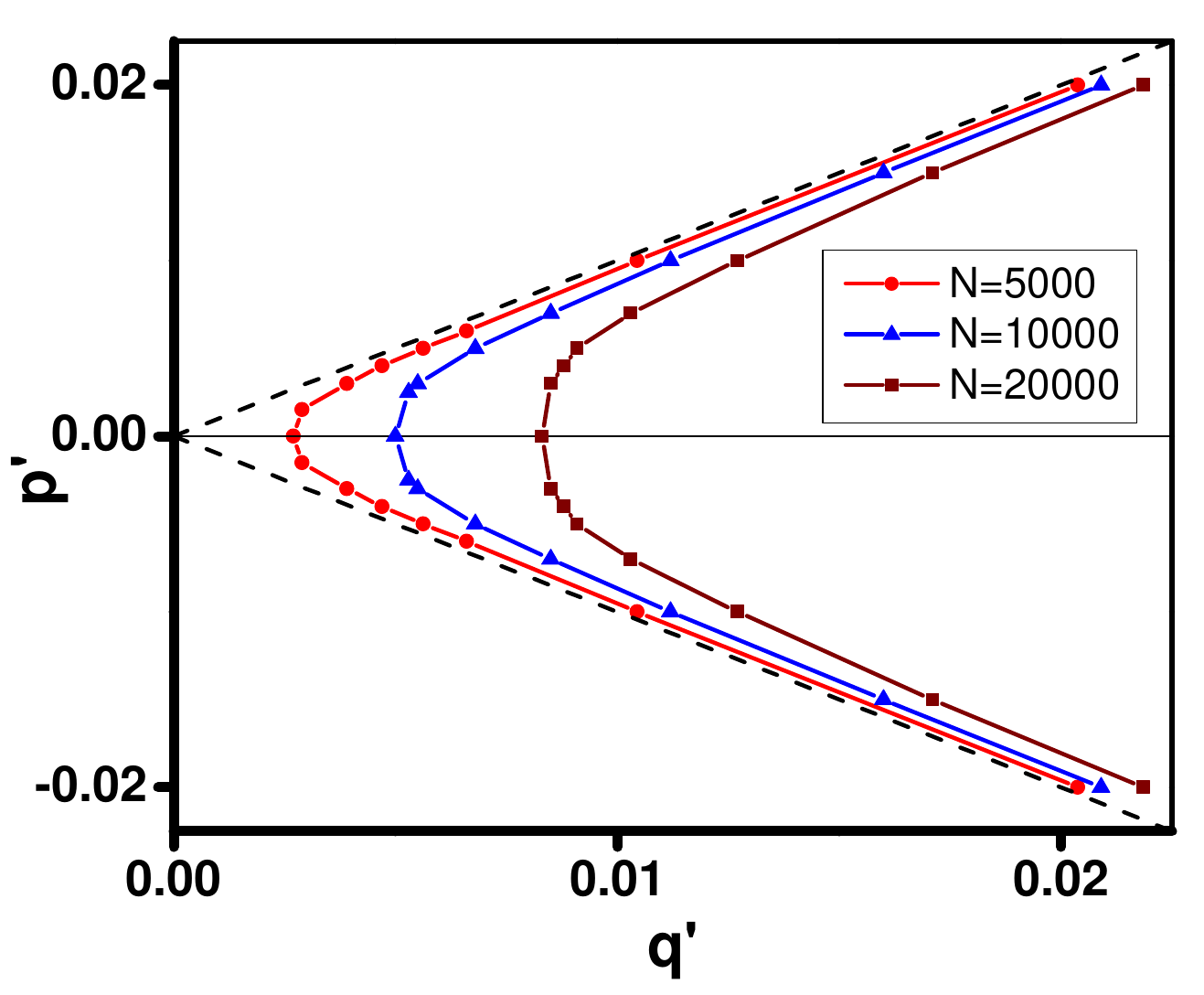}
	}
    \subfloat[Phase boundaries of the PM-polar phase transition after scaling with the factor $q_t$\label{subfig-2: phase_boundary_after_scaling}]{%
		\includegraphics[width=0.40\textwidth]{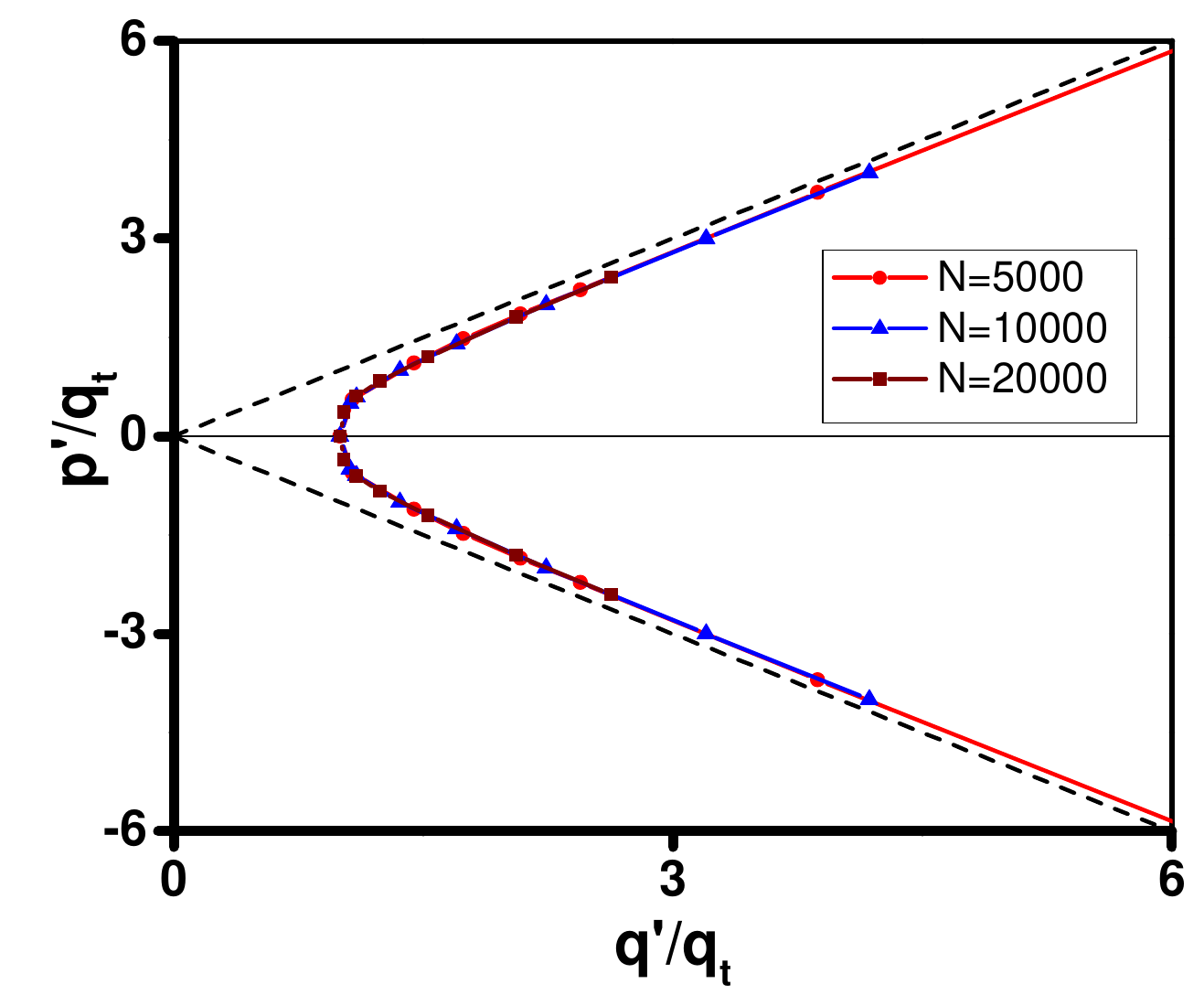}
	}	
	\caption{Subfig-(a): PM-polar phase transition boundary in p, q parameter space under harmonic confinement for 5000, 10000, and 20000 condensate particles (in red, blue, and wine markers). All the phase boundaries asymptotically follow $|p|=q$ line for large values of $p$ and $q$, similar to the homogeneous condensate, while if we increase the number of particles, the range of $q$ at $p=0$, for which the PM state becomes the ground state increases. Subfig-(b): The phase boundaries are plotted by scaling the quadratic and linear Zeeman terms with $q_t$, where $q_t$ is the quadratic Zeeman strength for which the PM-polar transition happens at $p=0$ for a particular N. In these scaled coordinates, all the phase boundaries approximately follow the equation, $(q'/q_t)^2-(p'/q_t)^2=1$, the equation of a hyperbola, similar to the homogeneous condensates. (color online)}
	\label{fig:trapped_phase_daigram_cleq0}
\end{figure*}
We first implement the VM for a condensate of N=5000 under a 3-D isotropic harmonic confinement of trapping frequency $\omega=2\pi\times 100\hspace{0.1cm} Hz$. The VM estimates total energy of the polar state and the PM state for different $p'$ and $q'$, a comparison of which would reveal the phase boundary for the trapped condensate. For homogeneous condensate, if we look at the energy difference between these two states for $p=0$ (Fig.\ref{fig:energy_PM_pol}\subref{subfig-1: hom_energy}), one can see that, the energy of the PM state is lower than the polar state for small positive values of $q$. As the strength of the quadratic Zeeman term is increased the energy difference reduces and at the transition point $q_t=2|c_1|n$ it vanishes. At this point, the number density of the $m=\pm 1$ projection vanishes (Eq.96 in \cite{KAWAGUCHI2012253}), hence the PM state ceases to exist. For the trapped condensate at $p=0$, one can estimate the $q_0$, where the peak density (number density at the center of the harmonic trap) of the $u_{\pm{1}}$ vanishes. Hence, in a similar manner, the PM state ceases to exist for $q>q_0$. The VM estimated energy difference between the PM and the polar state (Fig.\ref{fig:energy_PM_pol}\subref{subfig-2: VM_energy}) for $N=5000$ indicates that the phase transition happens at $q'_t=0.0027$, which is an order of magnitude lower than $q'_0\simeq0.0377$ beyond which the PM state ceases to exist under trapped conditions.   

\par
As is evident from Fig.\ref{fig:homogeneous}, for the homogeneous condensate, the constant number density is used in the scaling of $p$ and $q$. As a result, the whole phase diagram is universal with respect to number density variation for any homogeneous spin-1 condensate. In contrast, in the presence of confining potential, the number density varies over space and even the peak density (number density at the center of the trap) is different for different stationary states. In this case not the number density but the number of condensate particles are of importance.
\par
The VM can be employed for different values of $p'$, to estimate the $q'$ value for which the energy difference of the PM and the polar state vanishes. Thus, one can get the phase boundaries (Fig.\ref{fig:trapped_phase_daigram_cleq0}\subref{subfig-1: trapped_phase_daigram}) in the $q'$, $p'$ parameter space for a range of condensate particles. Note that, these phase transition boundaries are similar and asymptotically approach the $|p|=q$ line for large values of $q$. At the same value of the linear Zeeman term $p$, with an increase in the number of particles, the phase transition happens at a higher value of $q$. For example, at $p=0$, we previously mentioned that the phase transition happens at $q_t=0.0027$, which gets shifted to $q_t=0.005$ for N=10000 and $q_t=0.0083$ for N=20000.
\begin{figure}[htp]
    \centering
    \subfloat{%
		\includegraphics[width=0.40\textwidth]{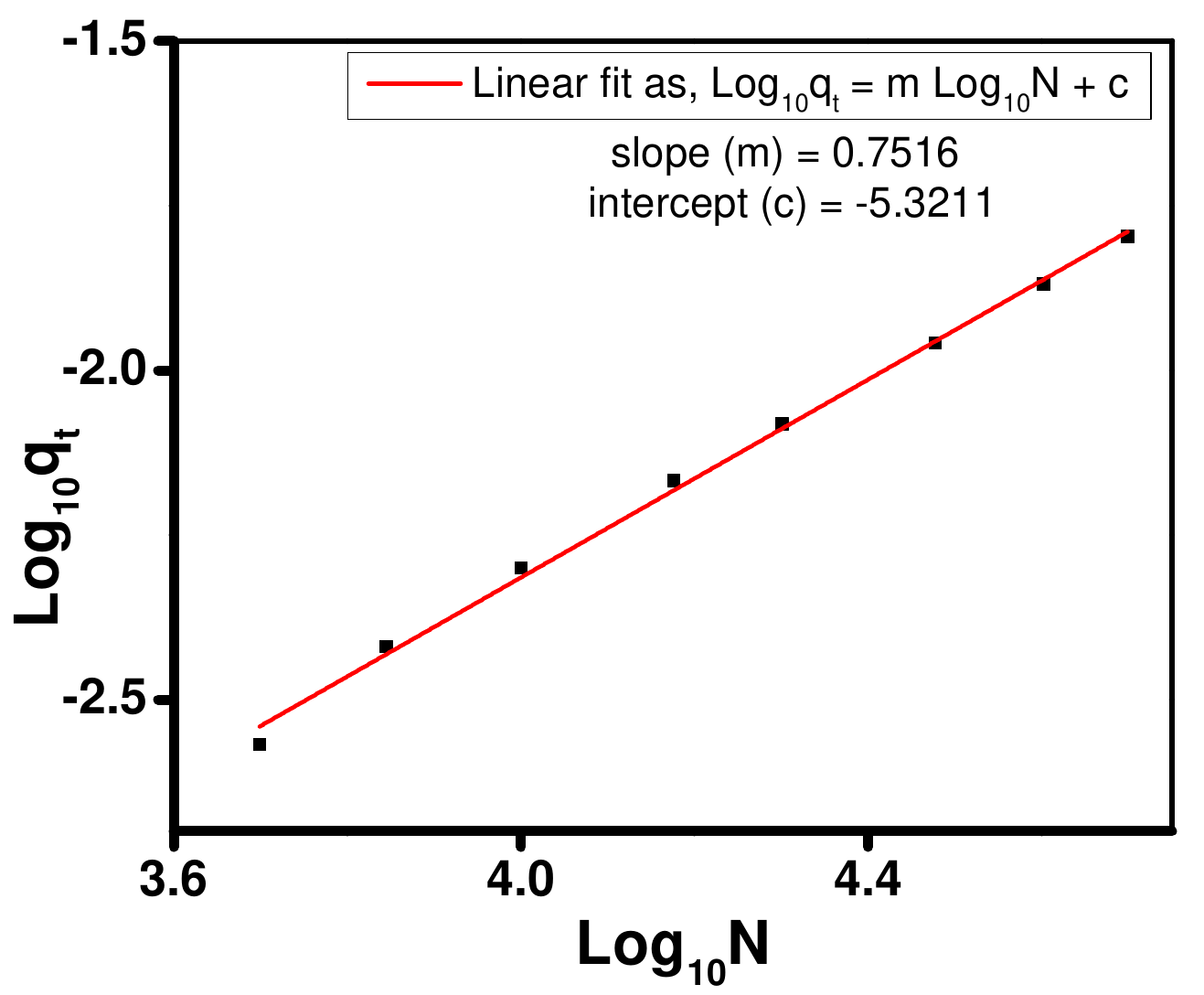}
	}
        
	\caption{The scaling factor $q_t$ depends on the number of condensate particles. Assuming a power function $q_t\propto N^m$, from the log-log plot, the slope corresponding to the exponent $m$ is obtained, i.e., $m\approx 0.75$. So, the scaling factor $q_t$ varies approximately as $q_t\propto N^{3/4}$.}
	\label{fig:scaling}
\end{figure}
\par
A natural query, therefore, would be whether there exists a scaling factor for trapped condensate which brings these phase boundaries for different numbers of condensate particles (Fig.\ref{fig:trapped_phase_daigram_cleq0}\subref{subfig-1: trapped_phase_daigram}) to the same plot. The asymptote of unit slope indicates, if we scale the $p'$ and $q'$ with $q_t(N)$, which is the $q$ value at $p=0$, where the transition happens for a particular N, then all these phase boundaries merge (Fig.\ref{fig:trapped_phase_daigram_cleq0}\subref{subfig-2: phase_boundary_after_scaling}) and approximately follow $(q'/q_t)^2-(p'/q_t)^2=1$, the equation of a hyperbola, similar to the homogeneous condensate. The scaling factor $q_t$ depends on the number of condensate particles roughly as $q_t\sim N^{3/4}$ (Fig.\ref{fig:scaling}).     
\par
In the absence of trapping, the PM-polar phase transition happens at a quadratic Zeeman term, $q_t=2|c_1|n$ at $p=0$. In that case, the number density is a constant. If we map the T-F approximated trapped condensate to the homogeneous counterpart by replacing the constant density with the T-F approximated average density, one can get to an estimation of the $q^{T-F}_t$ in the following way.
\par
The total number present in the condensate can be obtained by integrating the number density. In the T-F approximation,
\begin{equation}\label{eq:rev40}
    \int_0^R 4\pi \zeta^2 \Big[\mu'_{T-F}-\dfrac{1}{2}\zeta^2\Big] d\zeta \sim N,
\end{equation}
where $R$ is the T-F radius and $\mu'_{T-F}=R^2/2$. So, the T-F radius varies with N as, $R\sim N^{1/5}$, which leads to the volume $V\sim R^3\sim N^{3/5}$. As a result the average density in the T-F approximation $n_{avrg}=N/V\sim N^{2/5}$. If we replace the constant density with the T-F approximated average density, one can estimate $q_t^{T-F}\sim N^{2/5}$. In contrast, the multi-modal variational method indicates that this scaling factor is $q_t\sim N^{3/4}$ for the isotropic 3-D harmonic confinement. This result exposes the limitation of $T-F$ approximation in comparison with more accurate results even in large condensates.

\section{Discussion}
We have presented, in this paper, an accurate analytical description of the multi-component ground states of a harmonically trapped spin-1 condensate. Even in the so-called "T-F regime", where the overall density of the condensate is high enough to supposedly neglect the kinetic energy contribution, the T-F approximation can go wrong when applied to multi-component states. On the other hand, the SMA can, as well, be significantly inaccurate in handling such multi-component situations. This requires a general multi-modal treatment taking into consideration the kinetic energy term, which the VM provides. The VM correctly captures multi-component ground states because it can accurately estimate trapped density profiles even in the low-density regions. These tail parts of low densities are where the kinetic energy contribution is more significant than the interaction energy.
\par
Moreover, the VM can be easily implemented for 3-D harmonic trapping, where doing numerical simulation is well-known to be computationally expensive. Utilizing this advantage, we further explored the phase boundary between the phase-matched and the polar state under isotropic harmonic trapping. We have presented a detailed analysis of the shift of the phase boundary in the trapped case which, despite being significantly shifted from the homogeneous case, these boundaries bear clear qualitative correspondence with the homogeneous case. The scaling of these phase boundaries with the particle number of trapped condensates is found out. This scaling deviates significantly from that estimated from a T-F approximated trapped case equivalent to the homogeneous case.

\section{Acknowledgement}
PKK would like to thank the Council of Scientific and Industrial Research (CSIR), India for providing funding during this research. The support and the resources provided by PARAM Brahma Facility under the National Supercomputing Mission, Government of India at the Indian Institute of Science Education and Research; Pune are gratefully acknowledged.
\vspace{0.5cm}
\begin{flushleft}
\textbf{Data availability statement:} The datasets generated and analyzed during the current study are available from the corresponding author at a reasonable request.
\end{flushleft}
\bibliographystyle{apsrev4-1}
\bibliography{main} 	 
\appendix
\appendix
\section{Variational Method}
To work with the non-zero contribution of $p$ and $q$, we extend the variational method introduced in the article \cite{kanjilal_variational}. First, we will present the general method in a brief manner and then we will implement that for the PM state and the AF state which are the multi-component states of interest to this article.

\par

To implement the variational method, in presence of quasi-one-dimensional harmonic trapping, one needs to solve the GP equations Eq.\ref{eq:rev13}-\ref{eq:rev14} by getting rid of the kinetic energy terms which will yield the sub-component densities in the high-density region near the center of the trap, where the kinetic energy can be neglected in comparison to the interaction terms. The sub-component densities will get a functional form specific to different stationary states. Next, we assume that near the low-density region where the kinetic energy is of relevance and the interaction terms are very small, the mean fields can be described in terms of the first few lowest harmonic oscillator states. Thus the density distributions can be written as,
\begin{equation}\label{eq:app1}
u^{in}_{\pm 1,0}=g_{\pm 1,0}(\mu',\zeta), \hspace{0.3cm}\text{for}\quad|\zeta|\leq\zeta_{\pm 1,0}^{mat}
\end{equation}
\begin{equation}\label{eq:app2}
\begin{split}
u^{out}_{\pm 1,0}=(a_{\pm 1,0}+c_{\pm 1,0} |\zeta|+d_{\pm 1,0} &\zeta^2)exp\left(-\dfrac{\zeta^2}{b_{\pm 1,0}}\right)\\
&\hspace{0.0cm}\text{for}\quad|\zeta|\geq\zeta_{\pm 1,0}^{mat},
\end{split}
\end{equation}
where, $u^{in(out)}_{\pm 1,0}$ is the sub-component densities in the high-(low-) density region. Now we impose the condition that for each sub-component the low-density $\sqrt{u^{out}_{\pm 1,0}}$ and the high-density $\sqrt{u^{in}_{\pm 1,0}}$ expressions match at a point $\zeta^{mat}$. Not only do they match but their first three derivatives also match. These four constraints provide the four unknowns $a$, $b$, $c$, and $d$ for each sub-components in terms of the matching points and the parameter $\mu'$. Note that, imposing the matching condition up to three derivatives also gives a smooth profile of the corresponding kinetic energy.
\par
Once all the coefficients in Eq.\ref{eq:app2} are known, the sub-component density profile only depends on the parameter $\mu'$ and the matching points. The parameter $\mu'$ can be obtained as a function of the matching points from,
\begin{equation}
\begin{split}\label{eq:app3}
\sum_{m=-1}^1\Bigg[\int_0^{\zeta_m^{mat}}& u_m^{in}(\mu',\zeta)d\zeta\\
&+\int_{\zeta_m^{mat}}^{\infty}u_m^{out}(\mu',\zeta,\zeta_m^{mat})d\zeta\Bigg]=N,
\end{split}
\end{equation}
where N is the total number of condensate particles. Note that, one might expect that the right side should be $N/2$ as the integration is running in only one direction from the center of the trap, but it is N in the right side due to the Eq.\ref{eq:rev11}-\ref{eq:rev12} which we used to write the GP equation in non-dimensional form. From this Eq.\ref{eq:app3} the parameter $\mu'$ can be written as a function of the matching points for a particular N.
\par
Thus the sub-component number densities and hence the total energy of a stationary state (Eq.\ref{eq:rev10}) also becomes the function of the matching points only. From the minimization of the total energy in the parameter space of the matching points, one can determine the matching points as well as the total energy.

\subsection{Variational method for the PM state:}
For the PM state, all the sub-components are populated followed by the phase matching condition, i.e., the relative phase being $\theta_r=0$. 
One can solve the phase stationary equations (Eq.\ref{eq:rev13}-\ref{eq:rev14}) by ignoring the kinetic part to get the sub-component densities in the high-density region as,
\begin{equation}\label{eq:app4}
u^{in}_m=k_m\Bigg[\dfrac{\mu'_m-\zeta^2/2}{\lambda_0+\lambda_1}\Bigg],
\end{equation}
where, 
\begin{equation}\label{eq:app5}
k_1=\dfrac{(p'+q')^2}{4q'^2},\quad k_0=\dfrac{q'^2-p'^2}{2q'^2},\quad k_{-1}=\dfrac{(p'-q')^2}{4q'^2}
\end{equation}
and,
\begin{equation}\label{eq:app6}
\begin{split}
&\mu'_{\pm1}=\mu'_{eff}+(\lambda_0+\lambda_1)\dfrac{q'^2-p'^2}{2\lambda_1 q'},\\
&\mu'_0=\mu'_{eff}-(\lambda_0+\lambda_1)\dfrac{q'^2+p'^2}{2\lambda_1 q'},\\
&\mu'_{eff}=\mu'+\dfrac{p'^2-q'^2}{2q'}.
\end{split}
\end{equation}
Applying the four matching conditions mentioned earlier, the unknown coefficients in the low-density expression for each sub-component can be obtained as,
\begin{widetext}
\begin{equation}\label{eq:app7}
a_m=\dfrac{1}{-8\mu'_m+4\zeta_m^2}\Bigg(\mu'_m\Big(-56\mu'_m+70 \zeta_m^2+4\kappa_m\Big)-3\zeta_m^2\Big(14\zeta_m^2+\kappa_m-6\mu'_m\Big)\Bigg) exp\Bigg(\dfrac{12\zeta_m^2}{\kappa_m}\Bigg),
\end{equation}

\begin{equation}\label{eq:app8}
b_m=\dfrac{\kappa_m}{12},
\end{equation}

\begin{equation}\label{eq:app9}
c_m=\dfrac{48\zeta_m^3\Big(-12\mu'_m+6\zeta_m^2+\kappa_m\Big)}{\kappa_m^2} exp\Bigg(\dfrac{12\zeta_m^2}{\kappa_m}\Bigg),
\end{equation}
\begin{equation}\label{eq:app10}
\begin{split}
d_m=\dfrac{1}{2\zeta_m^2\Big(-2\mu'_m+\zeta_m^2\Big)}\Bigg(-6(\mu'_m)^2-\zeta_m^2&\Big(\kappa_m+13\zeta_m^2-6\mu'_m\Big)\\
&+\mu'_m\Big(14\zeta_m^2+\kappa_m-6\mu'_m\Big)\Bigg) exp\Bigg(\dfrac{12\zeta_m^2}{\kappa_m}\Bigg),
\end{split}
\end{equation}
\end{widetext}
where, $\zeta_m$ is an abbreviation for the matching point $\zeta_m^{mat}$ and,
\begin{equation}\label{eq:app11}
\begin{split}
\kappa_m=6\mu'_m &-9\zeta_m^2\\
&+\sqrt{36\mu_m^2-12\mu_m \zeta_m^2 +33 \zeta_m^4},
\end{split}
\end{equation}
given the sub-component densities in the low-density region are represented as,
\begin{equation}\label{eq:app12}
u^{out}_{m}=\dfrac{k_m}{\lambda_0+\lambda_1}\Big(a_m+c_m |\zeta|+d_m \zeta^2\Big)exp\Bigg(-\dfrac{\zeta^2}{b_m}\Bigg).
\end{equation}
Now applying Eq.\ref{eq:app3} one can find the parameter $\mu'$ (see $\mu'_{eff}$ expression in Eq.\ref{eq:app6}) for different values of $\zeta_m$. Note that, as $\mu'_{1}=\mu'_{-1}$ (see Eq.\ref{eq:app6}) the matching points are the same for these two components, i.e., $\zeta_{1}^{mat}=\zeta_{-1}^{mat}$. Thus, the total energy (Eq.\ref{eq:rev10}) for the PM state becomes only a function of the matching points.
\par
From a physical perspective, the high-density expressions of $u_0$ and $u_{\pm1}$ were found by getting rid of the kinetic terms in the Eq.\ref{eq:rev13}-\ref{eq:rev14}. So the high-density expressions given in Eq.\ref{eq:app4} are true as long as all the sub-components are in the high-density region. If we focus on the case presented in the article, for $p'=0.01$ and $q'=0.3$, near T-F radius the $u_{\pm 1}$ sub-component does not follow the high-density expressions. As a result, one can neglect the derivative term in Eq.\ref{eq:rev13} but cannot neglect the same in Eq.\ref{eq:rev14}. This precisely makes the multi-component stationary states beyond the reach of T-F approximation as long as the sub-components do not follow the same spatial distribution.
\par
Precisely because of the reason stated above one has to shift the focus toward the total density. So, instead of using the high-density expression $u^{in}_0$ we will use the total density expression,
\begin{equation}\label{eq:app13}
    u_{tot}^{in}=k_{tot}\Bigg[\dfrac{\mu'_{eff}-\zeta^2/2}{\lambda_0+\lambda_1}\Bigg],
\end{equation}
written in the same fashion as Eq.\ref{eq:app4}, where $k_{tot}=1$. Now instead of the $m=0$ component, the same expressions Eq.\ref{eq:app7}-\ref{eq:app10}, would also provide the total density expression in the low-density region ($u^{out}_{tot}$ in Eq.\ref{eq:app12}). The $u_0$ component can be found by subtracting the other two component densities from the total density.
\par
Thus, the total energy of the PM state only becomes a function of the two matching points, $\zeta_{tot}^{mat}$ and $\zeta_{\pm1}^{mat}$. Minimizing the total energy with respect to these two parameter variations one can get the matching points and the total energy itself. And as the matching points are found, the analytical density expressions are also obtained.
\par
\subsection{Variational method for the AF state:}
When the $u_1$ and the $u_{-1}$ components are populated (even though they are unequally populated for non-zero values of $p'$ (see Table 1)) the stationary state is referred to as the anti-ferromagnetic or in short, AF state. When both the sub-components are in the high-density regions, one can write the densities by neglecting the derivative terms in the phase equations as,
\begin{equation}\label{eq:app14}
    u^{in}_{\pm 1}=k_{\pm 1}\Bigg[\dfrac{\mu'_{\pm 1}-\zeta^2/2}{\lambda_0}\Bigg],
\end{equation}
where, $k_{\pm 1}=1/2$ and,
\begin{equation}\label{eq:app15}
    \mu'_{\pm 1}=\mu'-q'\pm \dfrac{\lambda_0}{\lambda_1}p'.
\end{equation}
The sub-component density $u_0$ is zero throughout.
\par
Note that, the high energy expressions are valid as long as both the sub-component density is high enough so that the derivative terms can be safely ignored. But for $p'=0.2$ and $q'=-0.5$ (the case we discussed), the $u_{-1}$ component has a lesser T-F radius than the other component. As a result of it near the T-F radius of the $u_{-1}$ component, the high-density expression of the $u_{1}$ component would be invalid, for the reasons stated earlier. So, we will take the total density and the $u_{-1}$ component to implement the variational method.
\par
In the high-density region, the total density can be written as,
\begin{equation}\label{eq:app16}
    u^{in}_{tot}=k_{tot}\Bigg[\dfrac{\mu'_{tot}-\zeta^2/2}{\lambda_0}\Bigg],
\end{equation}
where $k_{tot}=1$. Now one can write the low-density expressions as
\begin{equation}\label{eq:app17}
    u^{out}_{m}=\dfrac{k_{m}}{\lambda_0}\Big(a_{m}+c_{m} |\zeta|
    +d_{m} \zeta^2\Big)exp\Bigg(-\dfrac{\zeta^2}{b_{m}}\Bigg),
\end{equation}
where $u^{out}_{-1}$, and $u^{out}_{tot}$ represents the $m=-1$ sub-component and the total density in the low-density region of the trap. The coefficients will have the same expressions as Eq.\ref{eq:app7}-\ref{eq:app11}. Following the same method as explained earlier one can minimize the total energy corresponding to this stationary state in the parameter space of the matching points $\zeta_{tot}$ and $\zeta_{-1}$ which provides the analytical form of the full profile of the condensate in terms of the total density and the $u_{-1}$ component. By subtracting the $u_{-1}$ component from the total density profile one can get to the $u_{1}$ component.

\section{Variational method in 3-D isotropic harmonic confinement}
If we consider the condensate to be trapped by an isotropic 3-D harmonic confinement with trapping frequency $\omega$, one can write the GP equation in a non-dimensional form using Eq.\ref{eq:rev38}-\ref{eq:rev39}. In non-dimensional form, the GP equation can be written as, 
\begin{equation}\label{eq:app18}
    \begin{split}
        \bigg\{& -\dfrac{1}{2}\dfrac{1}{\zeta^2}\dfrac{d}{d\zeta}(\zeta^2\dfrac{d}{d\zeta})+\dfrac{1}{2}\zeta^2+ \lambda_0 u-\mu'\\
        &+\lambda_1\left(u_1+u_{-1}+2 \sqrt{u_{-1}u_1}\cos\theta_r\right) \bigg\} \sqrt{u_0}=0,
    \end{split}
\end{equation} 
\begin{equation}\label{eq:app19}
    \begin{split}
    \bigg\{& -\dfrac{1}{2}\dfrac{1}{\zeta^2}\dfrac{d}{d\zeta}(\zeta^2\dfrac{d}{d\zeta})+\dfrac{1}{2}\zeta^2+ \lambda_0 u-\mu' \pm \lambda_1\left(u_1-u_{-1}\right)\\
    &\pm p'+ q'\bigg\}\sqrt{u_{\pm1}}
    +\lambda_1 u_0\left(\sqrt{u_{\pm1}} +\sqrt{u_{\mp1}}\cos\theta_r\right)=0.
    \end{split}
\end{equation}
where, due to isotropy, we have only considered the radial part of the Laplacian in the spherical polar coordinate. 
\par
Similar to the 1-D situation one can estimate the high-density expressions by neglecting the Laplacian term for a stationary state, followed by the assumption Eq.\ref{eq:app12} serving as the number density profile in the low-density region. Here, $\zeta$ is the radial distance from the center of the trap. The unknown parameters in Eq.\ref{eq:app12} for the PM state under 3-D harmonic confinement, assume the same expressions given in Eq.\ref{eq:app7}-\ref{eq:app10}.
\par
The only difference with the 1-D situation is in the form of the relation that is used to estimate the parameter $\mu'$. Integrating the sub-component densities would provide the total number of condensate particles, which in the non-dimensional form can be written as (following the scaling Eq.\ref{eq:rev38}-\ref{eq:rev39})
\begin{equation}
\begin{split}\label{eq:app20}
\sum_{m=-1}^1\Bigg[\int_0^{\zeta_m^{mat}}& u_m^{in}(\mu',\zeta)\zeta^2 d\zeta\\
&+\int_{\zeta_m^{mat}}^{\infty}u_m^{out}(\mu',\zeta,\zeta_m^{mat})\zeta^2 d\zeta\Bigg]=\dfrac{N}{3}.
\end{split}
\end{equation}
From this equation, one can estimate $\mu'$ for the matching points $\zeta^{mat}_m$ for a condensate with N particles. 
\par
For the polar state, the implementation of the variational method is straightforward. Only the $u_0$ component is populated for the polar state, hence Eq.\ref{eq:app18} becomes trivial. The number density expression in the high-density region for this state can be obtained by neglecting the Laplacian term in Eq.\ref{eq:app19},
\begin{equation}\label{eq:app21}
    u^{in}_0\Big|_{polar}=\dfrac{\mu'-\dfrac{1}{2}\zeta^2}{\lambda_0}.
\end{equation}
The number density in the low-density region would follow,
\begin{equation}\label{eq:app21}
    u^{out}_0\Big|_{polar}=\dfrac{1}{\lambda_0}\Big(a_{0}+c_{0}\zeta
    +d_{0} \zeta^2\Big)exp\Bigg(-\dfrac{\zeta^2}{b_{0}}\Bigg),
\end{equation}
where the coefficients $a_0$, $b_0$, $c_0$, and $d_0$ follows the same expressions as Eq.\ref{eq:app7}-\ref{eq:app10}.
\par
For the polar state, only one component being present makes the variational method even easier to implement. Following the same method discussed earlier, one has to estimate the matching point $\zeta^{mat}_0$ by minimizing the total energy in the one-dimensional parameter space of $\zeta^{mat}_0$. Note that, the total energy of the polar state does not depend on the linear and quadratic Zeeman terms.
\end{document}